\documentclass[preprint,3p,authoryear,times]{elsarticle}

\journal{Transportation Research Part C}
\usepackage[fleqn]{amsmath}
\usepackage[]{amsfonts}
\usepackage[]{amsthm}
\usepackage[]{amssymb}
\usepackage{empheq}
\usepackage{lscape} 

\usepackage[mathlines]{lineno}

\usepackage[]{algorithm2e}
\usepackage[english]{babel}
\usepackage[T1]{fontenc}

\usepackage{hyperref} 
\usepackage{graphicx}
\usepackage{caption,subcaption}
\usepackage{tikz}
\usepackage{pstricks}
\usepackage{xstring}
\usepackage{multirow}
\usepackage{hhline}

\definecolor{darkblue}{rgb}{0,0,0.5}

\hypersetup{colorlinks=true,linkcolor=black,citecolor=darkblue,urlcolor=darkblue} 

\newcommand{\newref}[2]{\hyperref[#2]{#1~\ref*{#2}}} 

\usepackage{longtable}
\usepackage{pdflscape}
\usepackage{tabularx}
\usepackage[toc,page]{appendix}

\newtheorem{theorem}{Theorem}
\SetArgSty{textnormal}
\usepackage{float}

\begin{document}

\begin{frontmatter}

\title{Modal equilibrium of a tradable credit scheme with a trip-based MFD and logit-based decision-making}

\author[1]{Louis Balzer\corref{cor}} 
\cortext[cor]{Corresponding author.} 
\ead{louis.balzer@univ-eiffel.fr}
\author[1]{Ludovic Leclercq}

\address[1]{Univ Gustave Eiffel, Univ Lyon, ENTPE, LICIT, F-69675 Lyon, France}

\begin{sloppypar} 
\begin{abstract} 
The literature about tradable credit schemes (TCS) as a demand management system alleviating congestion flourished in the past decade. Most proposed formulations are based on static models and thus do not account for the congestion dynamics. This paper considers elastic demand and implements a TCS to foster modal shift by restricting the number of cars allowed in the network over the day. A trip-based Macroscopic Fundamental Diagram (MFD) model represents the traffic dynamics at the whole urban scale. We assume the users have different OD pairs and choose between driving their car or riding the transit following a logit model. We aim to compute the modal shares and credit price at equilibrium under TCS. The travel times are linearized with respect to the modal shares to improve the convergence. We then present a method to find the credit charge minimizing the total travel time alone or combined with the carbon emission. The proposed methodology is illustrated with a typical demand profile from 7:00 to 10:00 for Lyon Metropolis. We show that traffic dynamics and trip heterogeneity matter when deriving the modal equilibrium under a TCS. A method is described to compute the linearization of the travel times and compared against a classical descend method (MSA). The proposed linearization is a promising tool to circumvent the complexity of the implicit formulation of the trip-based MFD. Under an optimized TCS, the total travel time decreases by 17\% and the carbon emission by 45\% by increasing the PT share by 24 points.
\end{abstract}

\begin{keyword}
tradable credit scheme \sep trip-based MFD \sep user equilibrium \sep logit \sep mode choice.
\end{keyword}

\end{sloppypar} 
\end{frontmatter}



\begin{sloppypar} 
\section*{Highlights} 
\begin{itemize}
\item A TCS is introduced in the trip-based MFD framework with elastic demand.
\item The delay induced by one user on the others is analytically derived.
\item Modal equilibrium is computed via iterative quadratic programming.
\item Social optimal credit charge is determined per dichotomy.
\item The TCS reduces the total travel time and carbon emission.
\end{itemize}

\section{Introduction}\label{Sect1}

Congestion is a global issue as it increases travel times and vehicle emissions. It induces economic losses, harms the environment, and contributes to health problems. Congestion occurs when the number of vehicles exceeds the optimal capacity of the existing transportation facilities. Some network operators are using demand management strategies to decrease the number of cars in the urban network and increase the share of travelers riding public transport (PT). For example, some cities around the world, such as Singapore, London, and Stockholm, have implemented urban tolls to increase the travel costs of private vehicles downtown and foster public transport share (see \cite{Gu2018CongestionEvidence}). \cite{Levinson2010EquityReview} reviews the equity of road pricing. As pricing generates revenues for the regulator, these revenues need to be spent in a way the system is profitable for as many users as possible, by cutting taxes or improving the transit network, for example.
Alternative approaches are quantity-based regulations based on credits. The regulator set a cap on the number of vehicles allowed to drive on the network and let the users trade the driving rights between themselves after an initial allocation. Note that they can always choose another means of transportation other than a private car, free of credits. The concept of tradable driving rights for congestion management was first mentioned in \cite{Verhoef1997TradeableExternalities}. The driving rights take the form of a fixed quantity of credits distributed among the population. The advantages are that it is revenue-neutral since the trades occur exclusively between users, and the price is fixed by the offer and the demand. \cite{Yang2011ManagingCredits} was the first to formulate a Tradable Credit Scheme (TCS) for an urban network based on BPR (Bureau of Public Roads) (\cite{BureauofPublicRoads1964TrafficComputer.}) functions to characterize travel times. Their objective was to change the users' routes over the road network.

A recent review of TCS and tradable permit schemes can be found in \cite{Lessan2019Credit-Advances}.
Some contributions in the area of TCS for congestion management are further compared in Table~\ref{tab:TCS_review}. The vast majority represents the transportation network with links ruled by BPR-like functions or a single generic Vickrey's bottleneck (\cite{Vickrey1969CongestionInvestment}). With the BPR function, users can choose their paths in the network while overall link loading defines the average travel time: the peak hour is considered as a mean steady flow. With Vickrey's model, departure times are the users' degree of freedom. This model is dynamic, but it assumes all the travelers have the same travel distance and share a joint bottleneck. In both cases, some works consider elastic demand, i.e., accounting for transit or trip cancellation because of high travel costs.
Two types of decision models have been proposed: deterministic (DUE, for Deterministic User Equilibrium) or probabilistic (logit-based). In the first case, a user will always take the least expensive alternative. In the second case, a user will choose an alternative with a probability related to the cost of this alternative compared to the costs of other alternatives. The probability associated with an alternative can be seen as the fraction of the flow using this alternative.
In most proposed TCS, the price mechanism is not explicit and appears as a Lagrangian factor to enforce the credit cap in the equations. It corresponds to the inequality stating that the sum of the consumed credits cannot exceed the allocated credits. Furthermore, the price is non-zero if and only if equality holds. It is known as the market-clearing condition (MCC). In some works, the price is determined by an iterative mechanism (\cite{Ye2013ContinuousCredits}, \cite{Guo2019TradableConvergence}, \cite{Liu2020ManagingApproach} (submitted)). In \cite{Tian2015Day-to-DayCredits}, the focus stands with the credit market formulation, and the price is determined by a double auction market. In a double-auction market, buyers and sellers formulate respectively asks and bids with their own desired prices. The credit price is then determined by the offer and demand. Buyers offering higher prices and sellers asking lower prices are making trades, and the others do not.
\cite{Bao2019RegulatingExist} looks at the equilibrium for a TCS when users are free to choose their departure times. Two models are investigated: Vickrey's bottleneck and Chu's model (\cite{Chu1995EndogenousApproach}). The last one is close to the BPR function but considers the distribution of the departure times. The authors show that the credit price is not always unique at equilibrium in Vickrey's bottleneck case, but it is for Chu's model. The equilibrium of Vickrey's bottleneck with TCS can be computed analytically. A network of links ruled by BPR functions allows for an explicit formulation of the equilibrium and thus the use of optimization software. In the case of an unknown elastic demand, \cite{Wang2012Bisection-basedScheme} and \cite{Wang2014TrialCase} present respectively for a link and a network an iterative method to compute the charges per link to minimize the total travel time.
Most of the contributions on TCS minimize the total travel time. In \cite{Wang2020CombinationApproach}, the total vehicle emissions are also considered, and the Pareto front is drawn.

\begin{table}[h]
\caption{Some TCS investigations in the literature.}\label{tab:TCS_review}
\begin{tabular*}{\hsize}{@{\extracolsep{\fill}}lllll@{}}
Paper & Congestion model & Equilibrium & Elastic demand\\
\hline
\cite{Yang2011ManagingCredits} & BPR & DUE & yes\\
\cite{Ye2013ContinuousCredits} & BPR & Logit & no\\
\cite{Jia2016TrafficCommute} & Vickrey & DUE & no\\
\cite{Miralinaghi2019ManagingBehavior}& Vickrey & DUE & no\\
\cite{Nie2013ManagingScheme}& Vickrey & DUE & yes\\
\cite{Nie2012TransactionCredits}& Vickrey & DUE & yes\\
\cite{Nie2015AProblem}& Vickrey & DUE & no\\
\cite{Xiao2015TradableCommuters}& Vickrey & DUE & no\\
\cite{Xiao2013ManagingCredits}& Vickrey & DUE & no\\
\cite{Tian2013TradableUsers}& Vickrey & DUE & no\\
\cite{Liu2020ManagingApproach} & MFD & Logit & no\\
\cite{Tian2015Day-to-DayCredits} & Simulator (DynusT) & DUE & no\\
\cite{dePalma2018CongestionAnalysis}  & BPR & Logit & no\\
\cite{Guo2019TradableConvergence}  & BPR & DUE & yes\\
\cite{Miralinaghi2020DesignPerception}  & BPR & DUE & no\\
\cite{Miralinaghi2016Multi-periodSchemes}  & BPR & DUE & yes\\
\cite{Wang2012Bisection-basedScheme}& BPR (only one link) & DUE & yes (but function unknown)\\
\cite{Wang2012TradableUsers}& BPR & DUE & yes\\
\cite{Wang2014TrialCase}& BPR & DUE & yes (but function unknown)\\
\cite{Bao2019RegulatingExist} & Vickrey/Chu & DUE & no\\
\cite{Wang2020CombinationApproach} & BPR & DUE & yes\\
\\
\end{tabular*}
\end{table}

Static models do not account for the congestion dynamics in cities. Relationships between mean speed and density for urban networks were formulated in \cite{Godfrey1969TheNetwork} and \cite{Mahmassani1984InvestigationResults}. \cite{Daganzo2007UrbanApproaches} introduces the Macroscopic Fundamental Diagram (MFD) concept to formalize congestion dynamics while keeping the network still tractable at a large urban scale. The network outflow or speed depends on the accumulation. Its trip-based formulation, also known as speed-MFD or generalized bathtub (\cite{Mariotte2017MacroscopicModels}, \cite{Lamotte2018TheLengths}, \cite{Jin2020GeneralizedFlows}) permits to represent the heterogeneity of trip lengths.
However, analytical investigations to derive network equilibrium with the MFD are challenging as travel times can hardly be explicitly derived. \cite{Lamotte2018TheLengths} and \cite{Jin2020GeneralizedFlows} discuss the distribution of trip lengths and departure times. \cite{Liu2020ManagingApproach} (submitted) is the first work presenting a TCS using a more advanced dynamic framework to reproduce traveler behavior. Instead of having a single downstream bottleneck like in Vickrey's model, the authors represent the congestion dynamics with the MFD framework. The mean speed depends on the number of cars in the system, and the users have different trip lengths. The authors optimize a time-varying distance-based credit charge to make the users choose departure times to minimize the total travel time. A model-free optimization method is used (Bayesian optimization).

In this paper, we investigate the equilibrium distribution of the users between private cars and transit, considering a TCS and traffic dynamics with a trip-based MFD. We aim to investigate how TCS can foster PT when the demand is elastic, and user choices are based on the perceived costs of all alternatives. Most of the literature about TCS was about driving the users to choose optimal routes or departure times. Some works introduced elastic demand but without explicitly considering transit. Re-routing the drivers or spreading the demand over time mitigate the congestion and reduce the exhaust gas emission. However, switching modes can address other externalities, such as the scarcity of parking places or the ecological footprint of automotive fleets over their life cycles. It is an auspicious research direction and fits with a trip-based MFD framework as it considers the dynamics of the congestion. 

The proposed TCS is relatively simple: the credit charge is constant over time and independent of the travel distance. It is applied on a day-to-day basis: each evening, the users choose if they are taking the car or riding transit on the next morning depending on the travel times of each alternative and the credit price. They get an allocation of credits for free from the regulator and can trade them between each other using an ad hoc application. The credit price depends on the offer and demand. Its simplicity makes it realistic for real application and already provides substantial improvement of the travel conditions.

The users are assumed to have given trip lengths and departure times, and their only degree of freedom is their modal choice: car or transit. The analytical properties of the MFD are used in order to compute the gradient of the travel times with regard to the modal choices. This information is then used to derive the demand equilibrium. An application of this method optimizes the credit charge to reduce the total travel time and the total network emission. No assumptions are made about the credit price regulation mechanism. The credit price is thus treated as a variable along the modal shares to reach the modal equilibrium, which needs to satisfy the MCC, as in many contributions. 

Our work brings four contributions: the first is formulating a TCS for an MFD where the degrees of freedom are the modal choices. The second is quantifying the relationship between the travel times and the modal shares in a trip-based MFD framework. The third one is a method to compute the modal equilibrium of the TCS with MFD by using the linearization of the travel times to quantify the delay induced by one user on the other users. The fourth one is a simple method using the previous results to optimize the credit charge to improve social welfare.

The remainder of this paper is organized as follows. In \hyperref[sec::framework]{Sect.~\ref*{sec::framework}}, we present the framework. \hyperref[sec::equilibrium]{Sect.~\ref*{sec::equilibrium}} formulates the modal equilibrium and its computation. The quantification of the marginal delay induced by an user, i.e., the derivation of the travel times is presented in \hyperref[sec::derivation]{Sect.~\ref*{sec::derivation}}.
The credit charge optimization is discussed in \hyperref[sec::toll]{Sect.~\ref*{sec::toll}}. A numerical example is provided in \hyperref[sec::example]{Sect.~\ref*{sec::example}} for a realistic test case corresponding to the morning commute in Lyon Metropolis. \hyperref[sec::conclusion]{Sect.~\ref*{sec::conclusion}} concludes this paper. For convenience, the notations are summed up in \ref{sec::app_notations}.

\section{Methodological framework}\label{sec::framework}

The network is represented by a trip-based MFD framework considering the whole city as a single region (\cite{Mariotte2017MacroscopicModels}, \cite{Lamotte2018TheLengths}, \cite{Jin2020GeneralizedFlows}). The demand consists of $N$ groups describing different clusters of travelers, each cluster having the same OD pair and departure time. 

\subsection{Multi-modal traffic dynamics}

 Different OD pairs and/or departures times are associated with different groups. If different routes are considered for the same OD pair, each route is represented by a different group. All the users of the same group enter the network simultaneously (same departure time), follow the same route (same trip length), and have the same travel time for each mode. The only degree of freedom is the ratio of car users per group. The car ratio in the group $i$ is noted $x_i$, and the vector of car ratio of all the groups is $\mathbf{x} \in [0,1]^N$. The number of travelers in group $i$ is $\gamma_i$. It means, when the group $i$ is traveling, its contribution to the car accumulation is $\gamma_i x_i$.

In the general multi-modal case, the travel time $T_i^m$ of group $i$ with the mode $m$ is linked to the trip length $l_i^m$, the departure time $t_i$, and the speed $V_i^m$ by:
\begin{equation}
    l_i^m = \int_{t_i}^{t_i+T_i^m} V_i^m \text{d}t.
\end{equation}

In the MFD framework, $V_i^m$ is assumed to be the same for all users sharing the same mode at the same time. This speed corresponds to the multi-modal MFD curves, which usually depend on the accumulation of both cars $n_\text{car}$ and PT vehicles (usually buses) $n_\text{PT}$: $V_i^m = V_i^m(n_\text{car}(t), n_\text{PT}(t))$. Here, we further simplify this relationship by assuming that the PT offer and operations do not change and are defined by the actual functioning of the PT network. PT travel times change in time based on historical observations based on a typical day. That means we consider the changes in PT travel times related to the existing adaptation of timetables during the peak hour and usual traffic conditions. So, we retrieve PT travel times directly from existing timetables and usual PT travel times in the Lyon Metropolis network during peak hours with respect to a given OD pair: 
\begin{equation}
    V_i^\text{PT} = \frac{l_i^\text{PT}}{T_i^\text{PT}},
\end{equation}
where $l_i^\text{PT}$ and $T_i^\text{PT}$ are retrieved from the city planner and depend on the departure time and OD pair of the group $i$.

$V_i^\text{car}(t)$ should depends on $n^\text{car}(t)$ and $n^\text{PT}(t)$. Because we assume that the PT operation is the same every day, that means $n^\text{PT}(t)$ does not change over days. So, we can directly fit $V_i^\text{car}(t)$ as a function of $n^\text{car}(t)$ based on historical data, and this will consider the usual interactions between cars and PT over the network. SO, the car speed MFD reduces to a function of the accumulation of cars only. Thus the relationship for cars becomes:

\begin{equation}
    l_i^\text{car} = \int_{t_i}^{t_i+T_i^\text{car}} V_i^\text{car}(n^\text{car}(t)) \text{d}t.
\end{equation}
Note that from this point, we omit the super- and subscript 'car' to lighten the notation and the subscript $i$ in the speed as it is common for all the groups because they travel in the same reservoir.
Note that all the derivations in the paper are made with this simplified representation of the car MFD, but \ref{sec::app_3D} shows all required transformations to adjust the calculations to the original formulation of the multi-modal MFD (a.k.a. 3D MFD).

Since the network model is a trip-based MFD, the travel times can be calculated using the virtual traveler introduced in \cite{Lamotte2018TheLengths}. We follow the trajectory of a fictional traveler who enters the network at the origin of time. We define $t \mapsto f(t)$ the traveled distance of the virtual traveler as a function of the time, and $s \mapsto n(s)$ the accumulation as a function of the traveled distance. The travel time $T_i$ of group $i$ depends on its trip length $l_i$ and departure time $t_i$:
\begin{equation}\label{eq::time_integral}
    T_i = \int_{f(t_i)}^{f(t_i)+l_i} \frac{1}{V(n(s))} \text{d}s.
\end{equation}

We assume that the speed on the network is always greater than a minimal speed $V_0>0$ to avoid numerical issues. It means we assume the network never reaches a complete gridlock.

\subsection{Mode choice}

\begin{figure}[h]
\centerline{\includegraphics[width=0.75\textwidth]{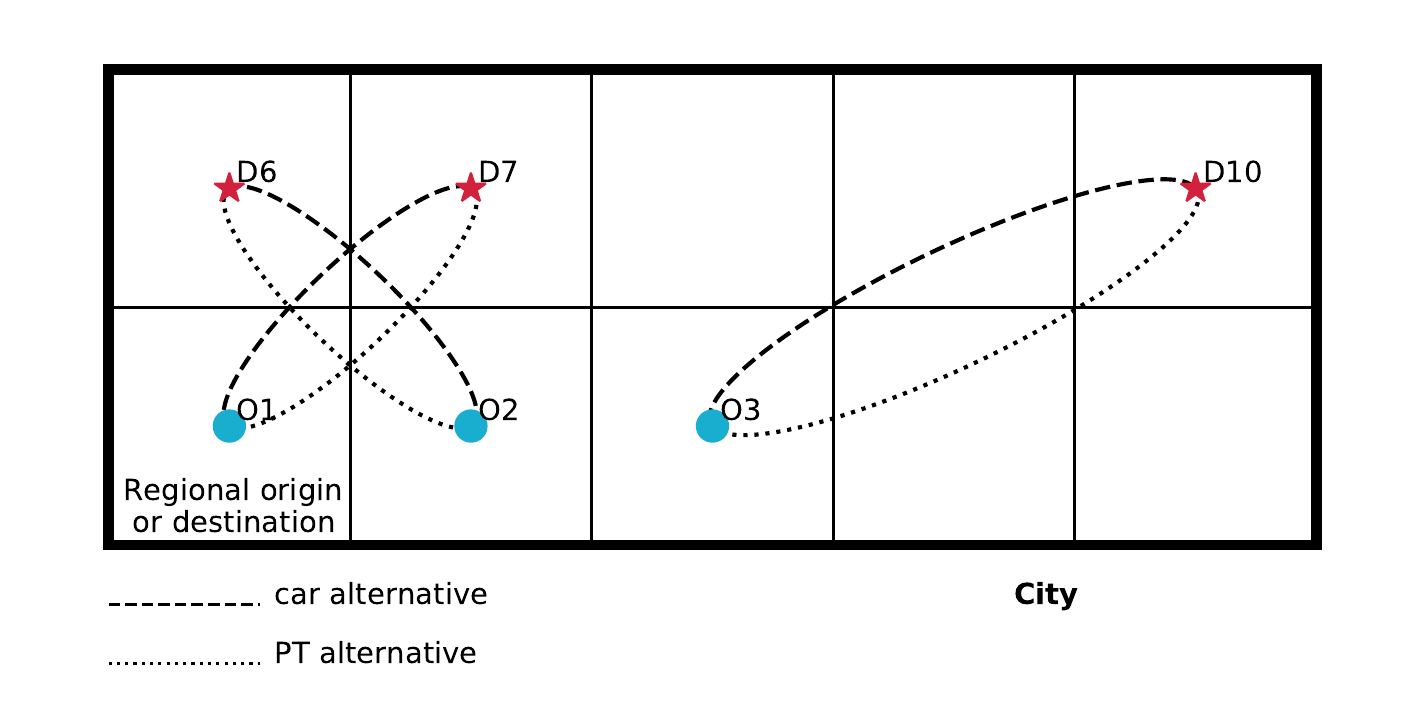}}
\caption{Each OD pair has car and PT alternatives.}
\label{fig:mode_choice}
\end{figure}

The users have two alternatives to complete their trips: private car or PT (See Fig.~\ref{fig:mode_choice}).

The travel costs of group $i$ for each mode are the monetary evaluation of the travel time plus the credit charge for the car:

\begin{equation}\label{eq::costs}
    \begin{cases}
    C_{i}^\text{car} &= \alpha T_i(\mathbf{x}) + (\tau-\kappa) p; \\
    C_{i}^\text{PT} &= \alpha T_{i}^\text{PT} - \kappa p,
    \end{cases}
\end{equation}
where $\alpha$ is the Value of Time (VoT), $\tau$ the credit charge, i.e., the number of credits one needs to take its car, $\kappa$ the allocation, i.e., the number of credits given by the regulator to each traveler for free, and $p$ the credit price, the money spent to buy one credit from another user or the money received after selling one credit to another user. Aside from the travel time difference, a PT user earns money by selling its credits since it does not need them. A car user spends money to purchase additional credits to pay the credit charge, since $\kappa<\tau$. Otherwise, the TCS is useless. The travelers are assumed homogeneous in the sense that they all have the same VoT. Considering heterogeneity is possible in this framework (with a VoT $\alpha_i$ specific to each group $i$) but not considered in this study.
A user taking the car has to spend $\tau$ credits. The group $i$ then spends in total $\gamma_i x_i \tau$ credits.

The number of cars allowed on the network per day is $\sum_i^N \gamma_i \frac{\kappa}{\tau}$. It is the parameter chosen by the control authority. In Sect.~\ref{sec::toll}, we assume the allocation $\kappa$ is fixed and we optimize the credit charge $\tau$. Optimizing the allocation under a fixed credit charge is equivalent, as only the ratio matters. 

The decision process follows a logit model. It assumes independent users' perceiving costs with an added error term following a Gumbel distribution. It is a well-established mode choice model that has a single parameter $\theta$. We adhere to the independence assumption for error between alternatives as costs for PT and cars do not depend on each other directly. The probability of group $i$ to drive a car given the modal shares $\mathbf{x}$ and its associated traffic conditions and the credit price is:

\begin{equation}\label{eq::logit}
    \psi_i(\mathbf{x}, p) = \frac{e^{- \theta C_{i}^\text{car} }}{e^{- \theta C_{i}^\text{car}} + e^{- \theta C_{i}^\text{PT}}}.
\end{equation}
Since each group represents several travelers, $\psi_i$ is used as the ratio of users in group $i$ willing to use the car. A similar approach is used in \cite{Ye2013ContinuousCredits}, where the logit is used not as a probability but as a  ratio of flows taking a particular path.

\begin{figure}[h]
\centerline{\includegraphics[width=0.49\textwidth]{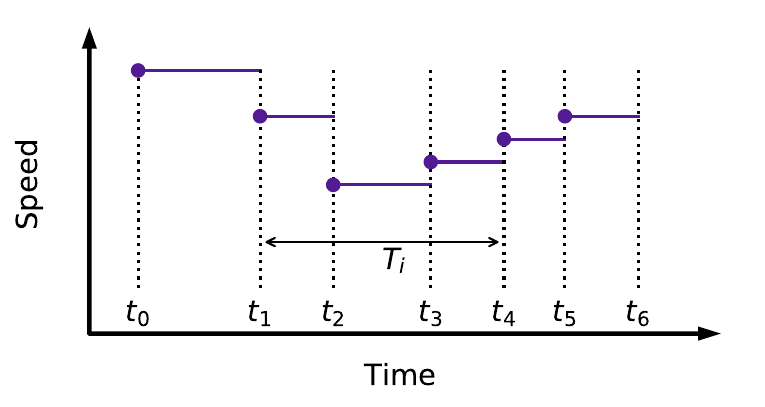}}
\caption{Decomposition of the travel time following events. In this illustration, group $i$ enters at the event 1 and exits at the event 4.}
\label{fig:time_decomposition}
\end{figure}

The travel time $T_i$ is computed by splitting the integral from Eq. \ref{eq::time_integral} every time a new event occurs, see Fig.~\ref{fig:time_decomposition}. An event is either the entry or the exit of a group in the network. Between two consecutive events, the accumulation does not change. Thus the speed is constant. We can then easily solve the integral as the terms under the small integrals are constant.
Let us note $e_{i,s}$ the event corresponding to the entry of group $i$ and ${e_{i,e}}$ the event relative to its exit. Then
\begin{align}\label{eq::time_sum}
    \begin{split}
    T_{i} &= \sum_{e=e_{i,s}+1}^{e_{i,e}} \int_{f(t_{e-1})}^{f(t_{e})} \frac{1}{V(n_{e-1})} \text{d}s\\
         &= \sum_{e=e_{i,s}+1}^{e_{i,e}} \frac{f(t_e)-f(t_{e-1})}
        {V(n_{e-1})}\\
        &= \sum_{e=e_{i,s}+1}^{e_{i,e}} T_e,
    \end{split}
\end{align}
with $T_e$ the time elapsed between the event $e-1$ and $e$, and $n_{e-1} = n(f(t_{e-1}))$.

\subsection{Network equilibrium}

Network equilibrium is reached when the actual mode shares are equal to the modal decisions given the same modal shares. The equilibrium is only implicitly defined as we need to know the modal shares to determine travel times, while mode shares calculations require travel times estimations. It is a classical fixed-point problem representing users who are satisfied with their assignments. It can be expressed as:

    \begin{subequations}
    \begin{align}
    \mathbf{x} &= \mathbf{\psi}(\mathbf{x}, p); \label{eq::SUE_eq}\\
    \tau \sum_{i=1}^N \gamma_i x_i &\leq \kappa \sum_{i=1}^N \gamma_i; \label{eq::SUE_TCS}\\
    x_i &\geq 0 \ \forall i; \label{eq::SUE_xl}\\
    x_i &\leq 1 \ \forall i; \label{eq::SUE_xh}\\
    p &\geq 0; \label{eq::SUE_pl}\\
    p \left(\sum_{i=1}^N \gamma_i (\kappa - \tau x_i)\right) &= 0. \label{eq::SUE_mcc}
    \end{align}
    \end{subequations}
Eq.~(\ref{eq::SUE_eq}) define the Stochastic User Equilibrium (SUE) under logit decision-making. Eq.~(\ref{eq::SUE_TCS}) is specific to the TCS: the number of consumed credits cannot exceed the overall allocation. Since the groups can trade credits between themselves, this constraint is at the system level and not the group level. Eq.~(\ref{eq::SUE_xl}), (\ref{eq::SUE_xh}) and (\ref{eq::SUE_pl}) delimit the admissible domain for the variables. There is no assumption on the credit price mechanism. It is a positive variable that has to be determined along with the modal shares. Eq.~(\ref{eq::SUE_mcc}) is the MCC as in \cite{Yang2011ManagingCredits}: the price is non-zero if and only if all the credits are consumed. 
TCS is a quantity-based demand management strategy. It means the number of trips by car is limited (Eq.~(\ref{eq::SUE_TCS})), but the price is not fixed. On the opposite, congestion pricing is a price-based strategy. The price is fixed, but the quantity is not limited. To change the proposed framework to congestion pricing, it is enough to remove Eq.~(\ref{eq::SUE_TCS}), (\ref{eq::SUE_pl}), and (\ref{eq::SUE_mcc}). The credit price $p$ would then be treated as a parameter.

\begin{theorem}
The proposed TCS admits at least one equilibrium state.
\end{theorem}

\begin{proof}
The proof of existence is inspired by \cite{Ye2013ContinuousCredits} and resorts to the fixed-point theorem.
First, let us define the following function $\Psi$, which represents a possible model for the system dynamics of the system:
\begin{equation}
    \Psi: (\mathbf{x}, p) \mapsto \left(\psi(\mathbf{x}, p), \left[p-\left(\sum_{i=1}^N\gamma_i(\kappa-\tau \psi_i\left(\mathbf{x}, p)\right)\right)\right]_+\right).
\end{equation}
The modal shares are updated following the logit-based decisions, and the price decreases if some credits are not used while always been positive. 
Let us show that $\Psi$ is continuous. The positive part function $[\cdot]_+$ is continuous. The accumulation between two consecutive events is the sum of the modal shares of the groups present on the network at that time, so the accumulation is continuous with regard to the modal shares. The speed $V$ is assumed continuous in the accumulation. The travel times $T_i$ are continuous in the speed $V$ (Eq. (\ref{eq::time_integral})). The modal choices $\psi_i$ are continuous in the travel times $T_i$ (Eq. (\ref{eq::costs}) and (\ref{eq::logit})), so the function $\Psi$ is continuous.

Let us find a compact convex $\Omega$ such as the image of $\Omega$ by $\psi$ stays in $\Omega$, i.e., $\Psi: \Omega \mapsto \Omega$. 
As the modal decision goes to zero when the price goes to infinity: 
$\forall \ i \in [1,N]$ and $\forall \ \mathbf{x}$, 
\begin{equation}
    0 \leq \psi_i(\mathbf{x}, p) \leq 1 - \frac{e^{- \theta \alpha T_{i}^\text{PT} }}{e^{- \theta \tau p}+ e^{- \theta \alpha T_{i}^\text{PT}}} \xrightarrow{p \rightarrow \infty} 0,
\end{equation}
we can find a price $p^* \geq 0$ satisfying $\sum_{i=1}^N\gamma_i(\kappa-\tau \psi_i(\mathbf{x},p)) \geq 0 \ \forall \ \mathbf{x}, \ p \geq p^*$. Let us set $p^+ = \max_{p\leq p^*}([p-\left(\sum_{i=1}^N\gamma_i(\kappa-\tau \psi_i(\mathbf{x},p))\right)]_+)$. Then setting $\Omega = [0,1]^N\times [0, p^+]$ works. 
By applying the fixed-point theorem to $\Psi$ on $\Omega$, we get a point $(\mathbf{x},p)$ satisfying $\mathbf{x} = \psi(\mathbf{x}, p)$ and $p=[p-\left(\sum_{i=1}^N\gamma_i(\kappa-\tau x_i)\right)]_+$. This couple $(\mathbf{x}, p)$ satisfies Eq. (\ref{eq::SUE_eq}-\ref{eq::SUE_mcc}), which proves the existence of an equilibrium.
\end{proof}

Proving the uniqueness of the solution is challenging because: (i) travel times have no explicit formulation, see Eq.~(\ref{eq::time_integral}) and (ii) the travel time of one group depends on the modal decisions of many other groups sharing the network at the same time. This coupling is the main difference with the previous contributions based on BPR-like functions, where the travel time on a link depends only on the number of vehicles on this link.
Here, we prove the uniqueness of the equilibrium under the condition that travel times increase strictly with the modal shares weighted by the number of travelers per group. Mathematically, it means that:
\begin{equation}\label{eq::monotony_tt_w_x}
    (\mathbf{T_1}-\mathbf{T_2})^T \cdot \gamma \cdot (\mathbf{x_1}-\mathbf{x_2})>0 \ \forall \ \mathbf{x_1} \ne \mathbf{x_2},
\end{equation}
with $\gamma$ being the matrix $N \times N$ with $\{\gamma_i, i \in [1,N]\}$ on the diagonal and zeros outside.
For a given test case, we can assess if this assumption is valid by randomly sampling multiple couples $(\mathbf{x_1}, \mathbf{x_2})$ and numerically verify through simulation that Eq.~\ref{eq::monotony_tt_w_x} always holds. It is not an absolute proof of uniqueness (which we believe is hardly possible because of the implicit nature and dependencies of $\mathbf{T}$), but, at least this provides a process to check uniqueness for any test case one like to study. Note that \ref{sec::app_uni} provides such a check for the numerical test case.

\begin{theorem}
If Eq.~(\ref{eq::monotony_tt_w_x}) holds, the equilibrium state is unique.
\end{theorem}

\begin{proof}
Let us take two equilibrium points [$\mathbf{x_1},p_1$] and [$\mathbf{x_2},p_2$].
Once again, the proof is inspired by \cite{Ye2013ContinuousCredits}. MCC and the credit cap tell us that:
\begin{equation}\label{eq::monotony_p_x}
\begin{split}
    &(p_1-p_2) \tau \sum_{i=1}^N \gamma_i \left( x_{1,i} - x_{2,i}\right)\\
    &= p_1 \sum_{i=1}^N \tau \gamma_i x_{1,i} -p_2 \sum_{i=1}^N \tau \gamma_i x_{1,i} - p_1 \sum_{i=1}^N \tau \gamma_i x_{2,i} + p_2 \sum_{i=1}^N \tau \gamma_i x_{2,i}\\
    &= p_1 \left(\sum_{i=1}^N \gamma_i \kappa - \sum_{i=1}^N \tau \gamma_i x_{2,i}\right) + p_2 \left(\sum_{i=1}^N \gamma_i \kappa - \sum_{i=1}^N \tau \gamma_i x_{1,i}\right)\\
    &\geq 0.
\end{split}
\end{equation}

By dividing the numerator and the denominator of the logit in Eq.~(\ref{eq::logit}) by $e^{-\theta \kappa p}$, for $i \ \in \ [1,N]$:
\begin{equation}
    \psi_i(\mathbf{x},p) = \frac{e^{-\theta (\alpha T_i(\mathbf{x}) + \tau p)}}{e^{-\theta (\alpha T_i(\mathbf{x}) + \tau p)}+e^{-\theta \alpha T_i^\text{PT}}},
\end{equation}
we remark that $\psi_i$ is decreasing with $\alpha T_i(\mathbf{x}) + \tau p$. Thus,
\begin{equation}\label{eq::monotony_cost_psi}
\begin{split}
    & \sum_{i=1}^N \gamma_i \left((\alpha T_i({x_1}) + \tau p_1)-(T_i({x_2}) + \tau p_2)\right) \left(\psi_i(\mathbf{x_1},p_1)-\psi_i(\mathbf{x_2},p_2)\right)\\
    & = \alpha (\mathbf{T(x_1)}-\mathbf{T(x_2)})^T \cdot \gamma \cdot (\mathbf{\psi_1}-\mathbf{\psi_2}) + \sum_{i=1}^N(p_1-p_2)\tau \gamma_i (\psi_{1,i} - \psi_{2,i})\\
    & = \alpha (\mathbf{T(x_1)}-\mathbf{T(x_2)})^T \cdot \gamma \cdot (\mathbf{x_1}-\mathbf{x_2}) + \tau (p_1-p_2) \sum_{i=1}^N \gamma_i (x_{1,i} - x_{2,i})\\
    &\leq 0.
\end{split}
\end{equation}
Eq.~(\ref{eq::monotony_tt_w_x}) makes the first term strictly positive for $\mathbf{x_1} \ne \mathbf{x_2}$. Using Eq.~(\ref{eq::monotony_p_x}), the second term is positive. It implies that $\mathbf{x_1 = x_2}$. As they are equilibrium points, $\mathbf{\psi_1 = \psi_2}$, and thus $p_1=p_2$ since the function $p \mapsto \psi(\mathbf{x_1}, p)$ is strictly decreasing. The equilibrium point is thus unique.
\end{proof}

In the scope of this work, we directly search the modal equilibrium without considering the time dynamics of the modal shares and the credit price. To assess the stability of the equilibrium points, we need to have a representation of the time (typically day-to-day) evolution of the modal shares and credit price when the modal assignment with TCS is not at equilibrium. To discuss the stability of the equilibrium, we provide a simple model to represent the time dynamics of the modal shares and credit price. The users update their modal shares to match their decisions, and the credit price is updated proportionally to the difference between credit supply and consumption. This representation is similar to the one used in \cite{Ye2013ContinuousCredits}. We prove the asymptotic stability of the equilibrium for our test case and different credit charges by calculating the eigenvalues of the Jacobian. See \ref{sec::app_stab} for more details.
\\

To conclude this section, let sum up the main assumptions of the modeling framework.
\begin{itemize}
    \item The trip lengths and departure times of the users are given for each OD pair.
    \item The travel times using PT only depend on departure time and OD pair.
    \item The travel times using the car depends on the time evolution of car accumulation, which results from all modal shares at the group level.
    \item The users' decisions follow a logit-based rule. They have the same VoT.
    \item The control authority uniformly distributes for free among all users a total quantity of credits equal to $\sum_{i=1}^N \gamma_i \kappa$. They then trade them between themselves.
    \item The credit price is strictly positive if and only if all credits are effectively used (MCC).
\end{itemize}

\section{Computing the modal equilibrium}\label{sec::equilibrium}

Contrarily to works based on Vickrey's bottlenecks and BPR functions, there is no implicit formulation of the SUE for a trip-based MFD formulation. We cannot directly transpose the existing methodology to calculate the equilibrium and have to develop a new one. This section presents the proposed workflow to find the modal equilibrium for a given credit charge, i.e., the number of credits needed to drive a car. It follows an iterative process based on the linearization of the equilibrium problem and the local resolution of a quadratic optimization problem (QP).

Let us start from an arbitrary mode choice vector $\mathbf{x_0}$ and a credit price $p_0$. The travel time for group $i$ is $T_{0,i}$ and its corresponding decision is $\psi_{0,i}$. The decision vector is noted $\mathbf{\psi_{0}}$. The car travel delay induced by the group $j$ on the group $i$ is noted $\nabla T_{i,j}$.

The vector of logit choices is linearized according to the change in the modal shares and credit price $\mathbf{\Delta \tilde x} = [\mathbf{\Delta x}; \Delta p]$:
\begin{equation}
    \mathbf{\psi} = \mathbf{\psi_{0} + \tilde \nabla \psi \cdot \Delta \tilde x + o(\Delta \tilde x)},
\end{equation}
where the operator $\tilde \nabla \cdot$ represents the gradient with respect to $\mathbf{\tilde x} = [\mathbf{x}; p]$. The gradient of the decision is defined by:
\begin{equation}\label{eq::logit_grad}
    \begin{cases}
        \psi_{0,i} &= \frac{e^{- \theta (\alpha T_{0,i} + (\tau-\kappa) p_0) }}{e^{- \theta (\alpha T_{0,i} + (\tau-\kappa) p_0) }+ e^{- \theta ( \alpha T_{i}^\text{PT} -\kappa p_0)}};\\
        \tilde \nabla \psi_{i,j} &= \psi_{0,i} (\psi_{0,i}-1) \theta \alpha \nabla T_{i,j};\\
        \tilde \nabla \psi_{i,N+1} &= \psi_{0,i} (\psi_{0,i}-1) \theta \tau,
    \end{cases} 
\end{equation}
for $i \in [1, N]$ and $j \in [1, N]$.
$\psi_{0,i}$ is the decision given the starting point, $\tilde \nabla \psi_{i,j}$ is the reaction of the group $i$ to an increase of the car share of a group $j$, and $\tilde \nabla \psi_{i,N+1}$ is the reaction of the group $i$ to an increase of the credit price.
We note that the sign of the gradient of the decision is opposed to the gradient of time. It confirms the general intuition that if the car travel time increases, the car will be less chosen. Similarly, if the credit price increases, the car will be less chosen as well.

The optimization process aims to find an equilibrium point, i.e., a point $\mathbf{\tilde x}$ satisfying $\mathbf{\psi = I_x \cdot \tilde x}$, with $\mathbf{I_x}$ the matrix of size $N \times (N+1)$ with 1 on the diagonal and 0 outside, such that $\mathbf{x = I_x \cdot \tilde x}$.

At the same time, the MCC should hold, i.e., the credit price is non-zero if and only if all the credits are consumed. We integrate the MCC in the cost function to not treat it as a quadratic hard constraint. It is numerically advantageous since all the constraints are then affine.

The objective function $J$ to minimize is defined as 

\begin{equation}
    J = \frac{1}{2}\left\lVert \mathbf{(\tilde \nabla \psi - I_x) \cdot \Delta \tilde x + \psi_0 - x_0} \right\lVert_2^2 + \eta p \frac{1}{\sum_{i=1}^N \gamma_i} \left(\sum_{i=1}^N \gamma_i (\kappa - \tau x_i)\right) ,
\end{equation}
with $\eta$ being the weight associated to the MCC.

\begin{figure}[h]
\centerline{\includegraphics[width=\textwidth]{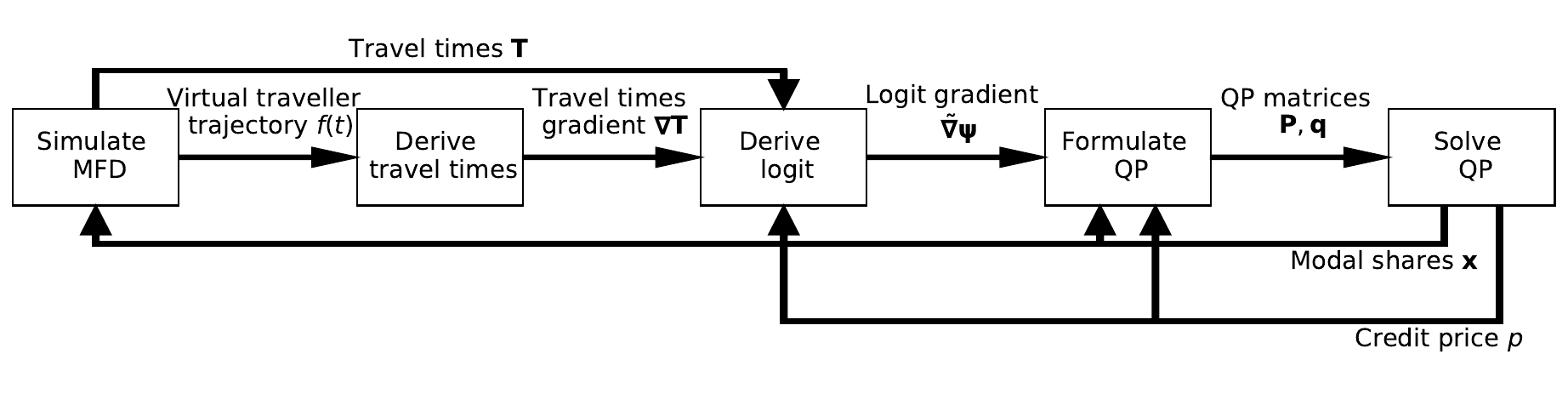}}
\caption{Flowchart of the search for the equilibrium.}
\label{fig:flow_eq}
\end{figure}

The optimization problem is formulated as a quadratic problem:

\begin{equation}\label{eq::QP}
    \frac{1}{2} \mathbf{\Delta \tilde x^T \cdot P \cdot \Delta \tilde x} + \mathbf{q \cdot \Delta \tilde x},
\end{equation}
by defining the symmetric matrix $\mathbf{P}$ and the vector $\mathbf{q}$ with
\begin{equation}
    \begin{cases}
        \mathbf{P} &= \mathbf{(\tilde \nabla \psi - I_x)^T \cdot (\tilde \nabla \psi - I_x)} + \eta \mathbf{I_p}; \\
        \mathbf{q} &= \mathbf{(\tilde \nabla \psi - I_x)^T \cdot (\psi_0 - x_0)} + \eta \mathbf{i_p},
    \end{cases} 
\end{equation}
where $\mathbf{I_p}$ is a symmetric matrix of size $(N+1)^2$ and $\mathbf{i_p}$ a vector of size $N+1$ defined by

\begin{equation} \label{eq::QP_mcc}
    \begin{cases}
        I_{p,i,N+1} = I_{p,N+1,i} &= - \frac{\gamma_i}{\sum_{j=1}^N \gamma_j} \tau  \ \text{for} \ i \in [1,N] \ \text{ and 0 elsewhere};\\
        i_{p,i} &= - \frac{\gamma_i}{\sum_{j=1}^N \gamma_j} \tau p_0  \ \text{for} \ i \in [1,N];\\
        i_{p,N+1} &= \frac{1}{\sum_{i=1}^N \gamma_i} \left(\sum_{i=1}^N \gamma_i (\kappa - \tau x_{0,i})\right).
    \end{cases} 
\end{equation}
The first terms of $\mathbf{P}$ and $\mathbf{q}$ stand for the modal equilibrium and the second ones stand for the MCC.
The constraints on the search space and on the credit consumption are then:
\begin{equation} \label{eq::QP_cons}
    \begin{cases}
        \Delta \tilde x_i &\leq \min(1-x_{0,i}, \epsilon_x) \ \text{for} \ i \in [1,N]\\
        \Delta \tilde x_i &\geq \max(-x_{0,i}, -\epsilon_x) \ \text{for} \ i \in [1,N]\\
        \Delta \tilde x_{N+1} &\leq \epsilon_p\\
        \Delta \tilde x_{N+1} &\geq \max(-x_{0,N+1},-\epsilon_p)\\
        \tau \sum_{i=1}^N \Delta \tilde x_i &\leq \kappa N - \tau \sum_{i=1}^N x_{0,i}.
    \end{cases} 
\end{equation}
As we linearize several terms around a starting point, we do not want to explore the entire solution space but only the local neighborhood to find a better local solution. This is why we introduce two thresholds $\epsilon_x$ and $\epsilon_p$ that represent the maximum variations allowed respectively for the modal shares and the credit price.
The new optimal solution $\mathbf{ \tilde x}$ is used as the new starting point for the next iteration, and a new QP is formulated and solved. It lasts until a given number of iterations occurred or the cost function $J$ has reached a satisfying precision. Fig.~\ref{fig:flow_eq} summarizes the workflow.

In the numerical application, the QP (Eq. (\ref{eq::QP}) and (\ref{eq::QP_cons})) is solved using the Python package CVXOPT (\cite{CVXOPT}).

We also implement the classical MSA algorithm as a benchmark (\cite{Sheffi1985UrbanMethods}).
For each iteration $k$, the modal shares are updated according to:
\begin{equation}
    \mathbf{x}_{[k+1]}=\mathbf{x}_{[k]} + \frac{1}{k}\left(\mathbf{\psi}(\mathbf{x}_{[k]},p)-\mathbf{x}_{[k]}\right).
\end{equation}

It is swift to compute and very generic. It can be used for several assignment problems: route, time, or mode choice. However, it does not deal with the credit price as it only updates the modal shares. It does not enforce the TCS conditions and, in particular, the MCC and the total credit cap as it cannot hurdle specific constraints. It means that by using the MSA to find the equilibrium, there is no guarantee that the number of car users does not exceed the limit imposed by the credit cap.
One could argue that some modifications can be implemented not to violate the TCS conditions. For the sake of simplicity, this path will not be investigated as it would add another level of iterations, and the MSA in this work only acts as a benchmark.

\section{Derivation of the travel times with respect to the modal shares}\label{sec::derivation}

Equilibrium calculation requires the computation of the variables $\nabla T_{i,j}$. 
The operator $\nabla \cdot$ is the gradient with respect to the modal shares $\mathbf{x}$.
The car travel time of group $i$ can be approximated by 
\begin{equation}
    T_i = T_{0,i} + \mathbf{\nabla T_i} \cdot \mathbf{\Delta x} + o(\mathbf{\Delta x}).
\end{equation}

The $\{\mathbf{\nabla T_i}, \forall \ i\}$, previously defined as the delay undergone by one group because of the others, can also be seen as the derivatives of the travel times with respect to the modal shares.

We aim to quantify how the groups' modal choices influence the travel times of another group \emph{a priori}, i.e., without running several simulations for testing every possible scenario or search direction. A similar idea was used by \cite{Simoni2015MarginalDiagram} for marginal cost-based pricing: the authors estimated the delay caused by one user for each time step to update the pricing scheme. The delay induced by one user on the others was quantified to change the urban toll for each period. The estimation was done \emph{a posteriori} as MATSim was used to simulate the network, and thus analytical derivations were limited.

By calculating the gradient of inter-event times $\nabla T_e$, we can find the gradient of travel time of group $i$ $\nabla T_i$ by summing the changes in each inter-event period during group $i$'s trip:
\begin{equation}\label{eq::time_grad_sum}
    \mathbf{\nabla T_{i}} = \sum_{e=e_s+1}^{e_e} \mathbf{\nabla T_e}.
\end{equation}

Let us name $l_e$ the distance traveled by the virtual traveler during $T_e$, i.e. between the events times $t_{e-1}$ and $t_e$ and let us note $V_e = V(n_{e-1})$ the speed during this period. The event-scale variables $T_e$, $t_e$, $l_e$ and $V_e$ satisfy the following equations:

\begin{equation}\label{eq::length_period_speed_event}
    l_e = T_e V_e,
\end{equation}
and
\begin{equation}\label{eq::time_period_event}
    T_e = t_e - t_{e-1}.
\end{equation}

\begin{figure}
\begin{subfigure}{.49\textwidth}
  \centering
  \includegraphics[width=.6\linewidth]{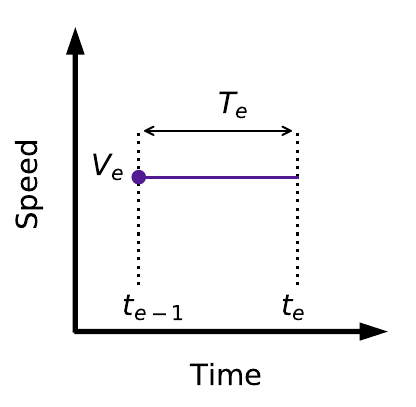}
  \caption{}
\end{subfigure}
\begin{subfigure}{.49\textwidth}
  \centering
  \includegraphics[width=.6\linewidth]{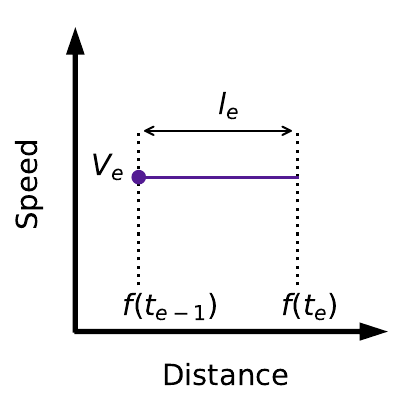}  
  \caption{}
\end{subfigure}
\caption{(a) Time- and (b) distance-based representations of the inter-event periods.}
\label{fig:event}
\end{figure}

These relationships can be seen in Fig.~\ref{fig:event}.

As the speed appears in the expression of $T_e$, it should be noted that its gradient $\mathbf{\nabla V_e}$ with respect to the modal shares $\mathbf{x}$ is expressed by:
\begin{equation}\label{eq::grad_speed}
    \begin{cases}
        \nabla V_{e,i} =  & \gamma_i \frac{\text{d}V}{\text{d}n}(n_{e-1}) \text{ if group $i$ is in the network between $e-1$ and $e$};\\
        & 0 \text{ otherwise}.
    \end{cases}
\end{equation}

We need to switch cases depending on the nature of the event $e-1$ and $e$. Note that as the departure times are assumed to be given, $\mathbf{\nabla t_e} = 0$ if $e$ is an entry. Furthermore, the trip length of a given group $i$ is constant too, so its gradient is zero $\mathbf{\nabla l_i} = 0$.

\begin{itemize}
    \item{Case I: $t_{e-1}$ and $t_e$ are both entries of groups in the network.}
    
    Since entry times are constant, by Eq. (\ref{eq::time_period_event}),
    \begin{equation}\label{eq::grad_entry_entry}
        \mathbf{\nabla T_e} = 0.
    \end{equation}
    
    \item{Case II: $t_{e-1}$ is an exit and $t_e$ is an entry.}
    
    Since $t_{e-1} = \sum_{g=1}^{e-1} \mathbf{\nabla T_g}$,
    \begin{equation}\label{eq::grad_exit_entry}
        \mathbf{\nabla T_e} = - \mathbf{\nabla t_{e-1}} = - \sum_{g=1}^{e-1} \mathbf{\nabla T_g}.
    \end{equation}
    
    \item{Case III: $t_{e}$ is the exit of a group $i$, i.e. $e_{i,e}=e$ ($t_{e-1}$ being an entry or an exit).}
    
    We decompose the trip length into the distance traveled between events, starting from the entry of $i$:
    \begin{equation}
        l_i = \sum_{g=e_{i,s}+1}^e l_g.
    \end{equation}
    By using Eq. (\ref{eq::length_period_speed_event}) and knowing that $l_i$ is constant, applying the gradient gives:
    \begin{equation}
        \sum_{g=e_{i,s}+1}^e \mathbf{\nabla T_g} V_g + T_g \mathbf{\nabla V_g} = 0.
    \end{equation}
    
    By calculating the gradient of inter-event period one after another in a time ascending manner, we can compute $\mathbf{\nabla T_e}$:
\begin{equation}\label{eq::grad_all_exit}
    \mathbf{\nabla T_e} = - \frac{1}{V_e} \left(T_e \mathbf{\nabla V_e} + \sum_{g=e_{i,s}+1}^{e-1} \mathbf{\nabla T_g} V_g + T_g \mathbf{\nabla V_g} \right).
\end{equation}
It is worth noticing the two parts of the gradient: a local contribution linked to the speed variation and the cumulative shift of the events. This shift is due to earlier or later completion of trips for groups ending their trips while group $i$ is in the network.
    
\end{itemize}

The gradient of the travel time is then computed following the algorithm \ref{algo::exact_grad}. The first loop addresses the events, and the inner ones focus on the groups. As there are $2N$ events (one entry and one exit per group), the number of operations to compute the gradient of the travel times $\mathbf{\nabla T}$ in one point $\mathbf{x_0}$ is $O(N^2)$, i.e., at most proportional to $N^2$.

\begin{algorithm}[H]\label{algo::exact_grad}
\SetAlgoLined
 \For{Each $\text{event } e \text{ in a time ascending manner}$}{
 \For{Each $\text{user } i \text{ present on the network at this time, i.e., } e_{i,s}<e\leq e_{i,e}$}
 {Compute the gradient of the speed $\nabla V_{e,i}$ according to Eq. \ref{eq::grad_speed}\;}
 Compute the marginal times $\mathbf{\nabla T_e}$ with Eq. \ref{eq::grad_entry_entry}, \ref{eq::grad_exit_entry} or \ref{eq::grad_all_exit}  depending on the types of the events $e-1$ and $e$\;
 \For{Each $\text{user } i \text{ present on the network at this time, i.e., } e_{i,s}<e\leq e_{i,e}$}
 {Add the contribution of this period $\mathbf{\nabla T_e}$ to the gradient  of the travel time $\mathbf{\nabla T_i}$ as in Eq.~\ref{eq::time_grad_sum}\;}
 }
 \caption{Computation of the gradient of the travel times relative to the modal choices.}
\end{algorithm}

\section{Optimization of the credit charge}\label{sec::toll}

Previously the credit charge was supposed given. However, the purpose of introducing a TCS is to improve the welfare of the society undergoing the externalities of network usage. 
In this study, we choose to minimize the total travel time only or combine total travel time and total network carbon emission by optimizing the credit charge level. Note that minimizing total carbon emission alone has a trivial optimal point: the credit charge level should be infinite, so everyone takes PT. This option has obvious drawbacks in terms of travel times (and acceptability), and this is why we better investigate a mixed objective function.

\subsection{Minimizing total travel time}

The first objective function is the total travel time $TTT$. It is the sum of the travel times per group and per mode, weighted by the corresponding modal ratio at equilibrium:
\begin{equation}
    TTT = \sum_{i=1}^N \gamma_i \left(x_i T_i + (1-x_i) T_{i}^\text{PT}\right).
\end{equation}

Let derive $TTT$ with respect to $\tau$ to determine its sign:

\begin{align}\label{eq::d_TTT_d_tau}
\begin{split}
    \frac{\text{d}TTT}{\text{d}\tau}
    &= \sum_{i=1}^N \gamma_i \frac{\text{d}x_i}{\text{d}\tau} (T_i-T_{i}^\text{PT}) + \gamma_i x_i \mathbf{\nabla T_i} \cdot \frac{\text{d}\mathbf{x}}{\text{d}\tau}.
\end{split}
\end{align}

Since we assume the network is always at equilibrium, the derivatives of $\mathbf{x}$ and $\mathbf{\psi}$ are equal. We remind that the modal choices $\mathbf{\psi}$ depend on the credit charge, credit price, and modal shares:

\begin{align}\label{eq::d_x_d_tau}
\begin{split}
    \frac{\text{d}x_i}{\text{d}\tau} =  \frac{\text{d}\psi_i}{\text{d}\tau} &= \frac{\partial \psi_i}{\partial \tau} + \frac{\partial \psi_i}{\partial p} \frac{\text{d} p}{\text{d} \tau} + \mathbf{ \nabla \psi_i \cdot \frac{\partial \psi}{\partial \tau}}.
\end{split}
\end{align}

Since the price mechanism is not explicit, $\frac{\text{d} p}{\text{d} \tau}$ would require numerical approximations.
For that, the solution space has to be sampled and the modal equilibrium calculated. We would then have the price for different credit charges and interpolate the derivative. This process is, however, costly and not fit for directly determining the optimal direction of the credit charge.
In order to circumvent the costly and prone-to-uncertainty estimation of the gradient of the price, a coarser but more robust and intuitive method is introduced. The general principle is to estimate the variations of the total travel time over an actual equilibrium for a given credit charge.

The changes in total travel time are the combined effect of improving the traffic conditions and the modal report. When the credit charge increases, the number of cars on the network decreases. The users still driving their cars benefit from better traffic conditions, and users shifting from car to PT usually experience an increase in travel time. The total travel time variation is estimated by: 
\begin{equation}
\begin{split}
    \Delta TTT&= N_c \Delta TT_c + \Delta N_c (TT_\text{c}^w - TT_\text{PT}^w),
\end{split}
\end{equation}
with $TT_\text{c} = \frac{\sum_{i=1}^N \gamma_i x_i T_i}{\sum_{i=1}^N \gamma_i x_i}$ the mean travel time per car and $N_c = \sum_{i=1}^{N} \gamma_i \frac{\kappa}{\tau}$ the number of car users supposing the credit cap constraint is active, i.e., all the credits are consumed. $TT_\text{c}^w = \frac{\sum_{i=1}^N \gamma_i w_i T_i}{\sum_{i=1}^N \gamma_i w_i}$ and $TT_\text{PT}^w = \frac{\sum_{i=1}^N \gamma_i w_i T_i^\text{PT}}{\sum_{i=1}^N \gamma_i w_i}$ are the mean travel time per car and per PT of users that are actually shifting from car to PT. The weights are the absolute values of the gradient of the logit $w_i = -\frac{\text{d} \psi_i}{\text{d} C_{i,\text{car}}}$. These weights give more importance to users prone to modal shift. By increasing the credit charge by a tiny quantity $\Delta \tau$, the $N_c$ car users will benefit from a reduction of their travel times by $\Delta TT_c$ and $\Delta N_c$ car users will be forced to switch to PT, increasing their travel time by $TT_\text{PT}^w - TT_\text{c}^w$.

By defining the typical accumulation on the network $\bar{n} = N_c \frac{TT_c}{T_\text{dept}}$, with $T_\text{dept}$ the time windows in which the departure times take place; the mean traveled distance by car, weighted by the modal shares $L_m = \frac{\sum_{i=1}^N \gamma_i x_i l_i}{\sum_{i=1}^N \gamma_i x_i}$; the mean speed over the whole simulation $\bar{V} = \frac{L_m}{TT_c}$ and the local slope of the speed $c$, such that $\Delta \bar{V} = - c \Delta \bar{n}$, we can derive the travel time variation of car users $N_c \Delta TT_c = L_m c \frac{1}{\bar{V}^2} \Delta \bar{n}$. The increase of the total travel time due to the modal shift is $(TT_\text{PT}^w - TT_\text{c}^w) \sum_{i=1}^{N} \gamma_i \frac{\kappa}{\tau^2} \Delta \tau$ and the decrease due to the improvement of the travel condition is $L_m c \frac{1}{\bar{V}^2} \bar{n} \sum_{i=1}^{N} \gamma_i \frac{\kappa}{\tau^2} \Delta \tau$.
Thus, the global variation of the total travel time becomes

\begin{equation}
\begin{split}
    \Delta TTT&= -L_m c \frac{1}{\bar{V}^2} \bar{n} \sum_{i=1}^{N} \gamma_i \frac{\kappa}{\tau^2} \Delta \tau + (TT_\text{PT}^w - TT_\text{c}^w) \sum_{i=1}^{N} \gamma_i \frac{\kappa}{\tau^2} \Delta \tau \\
    &= \left(-L_m c \frac{1}{\bar{V}^2} \bar{n} - TT_\text{c}^w + TT_\text{PT}^w \right) \sum_{i=1}^{N} \gamma_i \frac{\kappa}{\tau^2} \Delta \tau
\end{split}
\end{equation}

Thus the gradient of the total travel time can be approximated by:
\begin{equation}
    \frac{\text{d}TTT}{\text{d}\tau} \approx \left(-L_m c \frac{1}{\bar{V}^2} \bar{n} - TT_\text{c}^w + TT_\text{PT}^w \right) \sum_{i=1}^{N} \gamma_i \frac{\kappa}{\tau^2}
\end{equation}

\subsection{Minimizing the total network emission}

The total network emission of carbon dioxide is quantified using a macroscopic emission model COPERT IV for passenger cars (\cite{Ntziachristos2009COPERT:Model}). It quantifies the impact of network usage on global warming. It is also a proxy for fuel consumption. The PT part in emission is supposed constant because we assume that the PT operations are unchanged (same number of vehicles and timetables). A straightforward extension would be to correlate the emission to the change in PT operation to accommodate the demand. However, the contribution compared to personal cars is much lower, so this would change neither our conclusion nor the methodology. Only personal cars emission are considered in this work.

In \cite{Ingole2020MinimizingGating}, the authors estimate the emission of \ensuremath{\mathrm{CO_2}} and \ensuremath{\mathrm{NO_x}} in an accumulation-based MFD framework. 
For a given time period, emissions are the product of the total travel distance by all vehicles multiplied by the emission factor. The emission factor depends only on the mean speed. The total travel distance according to Edie's definition between two consecutive events is $n_e T_e V_e$.
The total carbon dioxide emission $E$ is estimated by summing the contributions from all the inter-event periods:
\begin{equation}
    E = \sum_{e=1}^{2 N} n_e T_e V_e E_\text{dist} (V_e),
\end{equation}
where $V \mapsto E_\text{dist}(V)$ is the emission model giving the emission per distance as a function of the mean speed.

The emission function representative for a French typical vehicle fleet is represented by the fourth-order polynomial from \cite{Lejri2018AccountingScale}, see Table~\ref{tab:emission_coeff} for the coefficient values:
\begin{equation}\label{eq::emission_dist}
\begin{split}
    E_\text{dist}(V) =& c_{1} V^4 + c_{2} V^3 + (c_{3}+2 c_{1} c_{0}^2) V^2
    + (c_{4}+c_{2} c_0^2) V + (c_{5}+\frac{c_{3}}{3} c_{0}^2+\frac{c_{1}}{5} c_{0}^4).
\end{split}
\end{equation}

\begin{table}[h]
\caption{Parameters for \ensuremath{\mathrm{CO_2}} emission curve for passenger cars. These numerical values are for speeds in km/h and emissions in g/km.}\label{tab:emission_coeff}
\centering
\begin{tabular}{cc}
Coefficient & Value \\
\hline
$c_{0}$ & 12.5 \\
$c_{1}$ & 1.304$\times 10^{-5}$ \\
$c_{2}$ & -0.003269 \\
$c_{3}$ & 0.3103 \\
$c_{4}$ & -13.52 \\
$c_{5}$ & 371.4 \\
\end{tabular}
\end{table}

As for the total travel time, a coarse but robust estimation of the variation of the emission is calculated to avoid requiring numerical approximations of the price gradient with respect to the credit charge. As before, the changes in the network emission come from the modal report (total traveled distance changes) and the improvement of the traffic conditions (mean speed changes). As the credit charge increases, the total emission decreases on one side because fewer users are taking their car and the total travel distance decreases. On the other side, the emission per distance decreases because the mean speed globally increases. It means:

\begin{equation}
\begin{split}
    \Delta E &= \Delta L_\text{tot} E_\text{dist}(\bar{V}) + L_\text{tot} \frac{d E_\text{dist}}{d V}(\bar{V}) \Delta V\\
    &=\Delta N_c L_m^w E_\text{dist}(\bar{V}) - L_\text{tot} \frac{d E_\text{dist}}{d V}(\bar{V}) c \Delta \bar{n},
 \end{split}
\end{equation}   
with $L_\text{tot} = \sum_{i=1}^N \gamma_i x_i l_i$ the total traveled distance of all the cars. It is equal to $\sum_{e=1}^{2 N} n_e T_e V_e$. $L_m^w = \frac{\sum_{i=1}^N \gamma_i w_i l_i}{\sum_{i=1}^N \gamma_i w_i}$ is the mean travel distance by car of users shifting to PT. It is weighted by the absolute values of the gradient of the logit. When the credit charge increases by a tiny $\Delta \tau$, $\Delta N_c$ are forced to shift to PT and thus the total traveled distance per car decreases by $\Delta N_c L_m^w$. On parallel, as the typical accumulation decreases by $\Delta \bar{n}$, the traffic conditions are improved, and the carbon emission per distance decreases by $-\frac{d E_\text{dist}}{d V}(\bar{V}) c \Delta \bar{n}$. 
The decrease of the carbon emission due to modal report is $L_m^w E_\text{dist}(\bar{V}) \sum_{i=1}^{N} \gamma_i \frac{\kappa}{\tau^2} \Delta \tau$ and the decrease due to the better traffic condition is $-L_\text{tot} \frac{d E_\text{dist}}{d V}(\bar{V}) c \bar{n} \frac{1}{\tau} \Delta \tau$.
The global variation of the carbon emission becomes:
\begin{equation}
\begin{split}
    \Delta E &=-L_m^w E_\text{dist}(\bar{V}) \sum_{i=1}^{N} \gamma_i \frac{\kappa}{\tau^2} \Delta \tau + L_\text{tot} \frac{d E_\text{dist}}{d V}(\bar{V}) c \bar{n} \frac{1}{\tau} \Delta \tau\\
    &=\left( -L_m^w E_\text{dist}(\bar{V}) \sum_{i=1}^{N} \gamma_i \frac{\kappa}{\tau} + L_\text{tot} \frac{d E_\text{dist}}{d V}(\bar{V}) c \bar{n} \right) \frac{1}{\tau} \Delta \tau.
\end{split}
\end{equation}

The gradient of the total network emission is then approximated using
\begin{equation}
    \frac{\text{d}E}{\text{d}\tau} \approx \left( -L_m^w E_\text{dist}(\bar{V}) \sum_{i=1}^{N} \gamma_i \frac{\kappa}{\tau} + L_\text{tot} \frac{d E_\text{dist}}{d V}(\bar{V}) c \bar{n} \right) \frac{1}{\tau}.
\end{equation}
As the mean speed decreases with the network accumulation ($c\geq0$) and the emission per distance decreases with speeds on the typical range for urban network $\left(\frac{d E_\text{dist}}{d V}(\bar{V})\leq0\right)$, the gradient of the total network emission is always negative.

Note that this estimation and gradient method can be applied to other pollutants such as \ensuremath{\mathrm{NO_x}} or PM as the emission functions are similar. Emission curves with coefficient values can be found in \cite{Lejri2018AccountingScale}.

\subsection{Mixed objective function considering both emissions and travel times}

The objective function is the monetary evaluation of the total travel time and network emission. It is chosen as $\alpha TTT+\Gamma P_\text{carbon} E$. $P_\text{carbon}$ is the price of the carbon per weight and $\Gamma$ is the coefficient associated to the \ensuremath{\mathrm{CO_2}} emission. It is used to compensate for the difference in the order of magnitude between the total \ensuremath{\mathrm{CO_2}} emission and the total travel time.

Once again, since the trip-based MFD relies on an implicit formulation of the travel times, the optimal credit charge will not be explicitly given. However, we can compute an approximation of the derivative of the objective function.
We propose to solve this minimization problem by dichotomy: the search domain is halved at every step by looking at the sign of the derivative $\alpha \frac{\text{d}TTT}{\text{d}\tau}+\Gamma P_\text{carbon} \frac{\text{d}E}{\text{d}\tau}$. 
A lower and higher bounds are chosen at the initialization, and the first credit charge is the average. The recursive process goes through the following steps:
\begin{itemize}
    \item The equilibrium is computed;
    \item The approximation of the gradient of the objective function with respect to the credit charge is computed;
    \item If negative, the new credit charge is the average of the previous credit charge and higher bound. The lower bound takes the value of the previous credit charge. If positive,  the new credit charge is the average of the previous credit charge and the lower bound. The higher bound takes the value of the previous credit charge;
    \item When the higher and lower bounds are equal, the process stops as a local minimum has been found.
\end{itemize}

As the credit charge value is taken as an integer, the process always ends in a finite number of steps.
The same process is used when minimizing $TTT$ only. Repeating the process with different starting points mitigates the risk of finding only a local minimum.

\section{Numerical example}\label{sec::example}
In order to illustrate the proposed method, we design a large-scale scenario using representative data from a regular morning peak hour in Lyon Metropolis.

\subsection{Case study}

\begin{figure}
\begin{subfigure}{.59\textwidth}
  \centering
  \includegraphics[width=.95\linewidth]{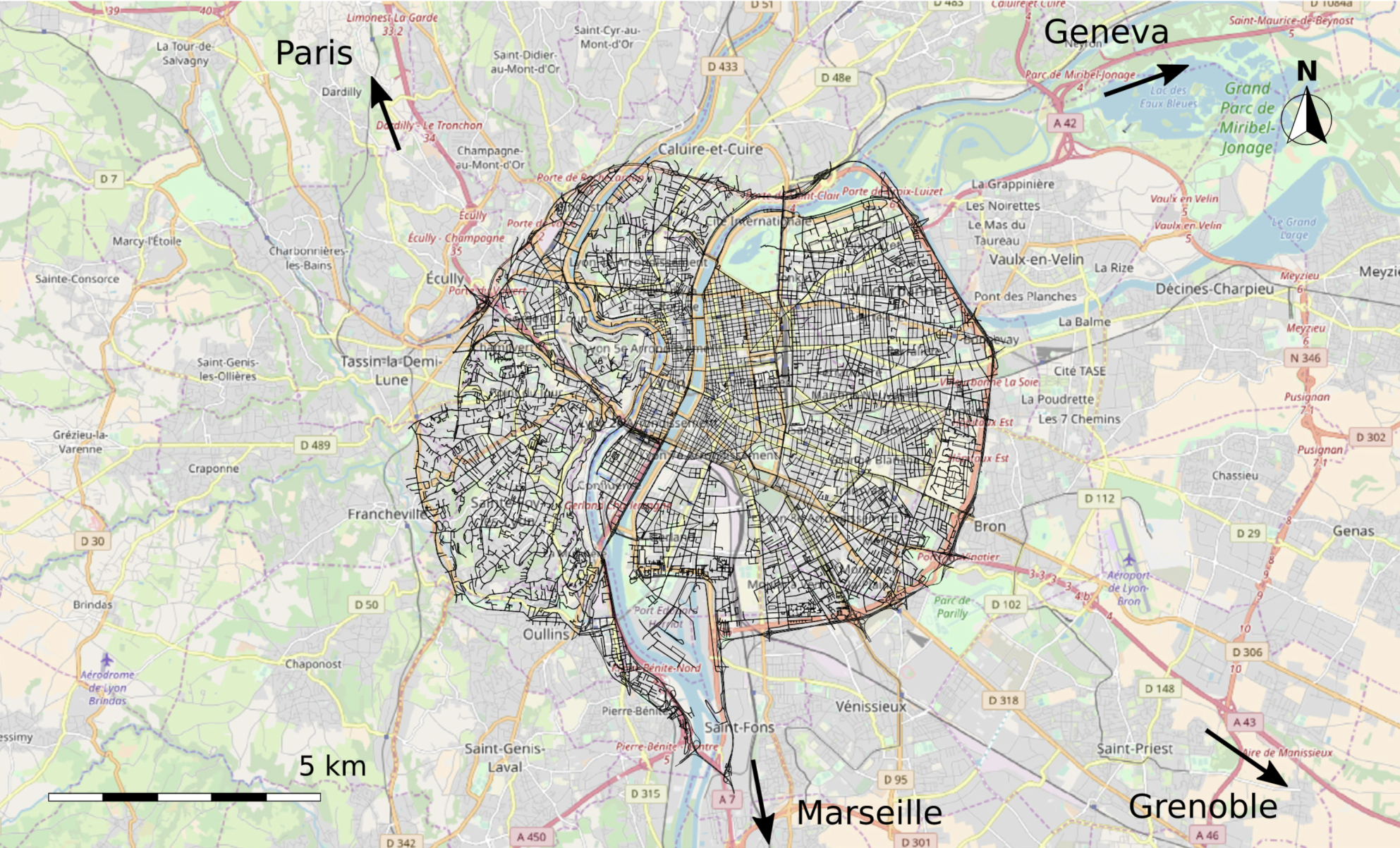}  
  \caption{}
\end{subfigure}
\begin{subfigure}{.4\textwidth}
  \centering
  \includegraphics[width=.95\linewidth]{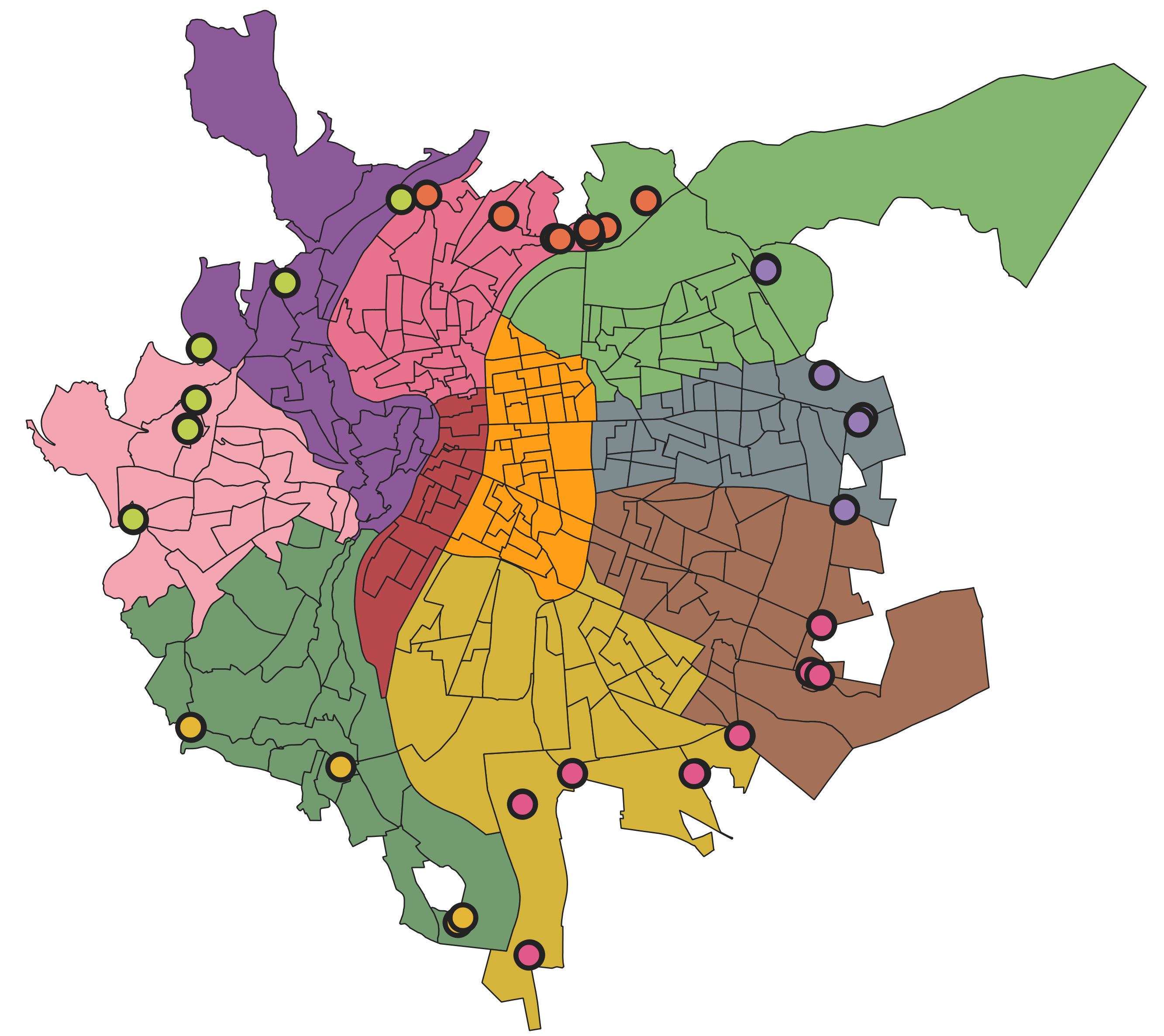}  
  \caption{}
\end{subfigure}
\caption{(a) The urban area under consideration (\cite{Mariotte2020CalibrationLyon}, \copyright OpenStreetMap); (b) The IRIS areas merged in 10 regions and the access points merged in five boundaries (circles).}
\label{fig:network}
\end{figure}

We use the network of Lyon Metropolis to calculate the travel times. The individual travelers are gathered into groups departing from an identical region at the same time and traveling to another common region. 
The MFD for the whole region has been experimentally determined in a previous study (\cite{Mariotte2020CalibrationLyon}). The demand is based on IRIS areas, which are French administrative areas with between 1~800 and 5~000 inhabitants. We regroup the OD pairs into a city partition of 10 regions to massify the demand and define relevant groups departing simultaneously. Furthermore, the perimeter is split into five regions to characterize trips starting or ending outside of Lyon Metropolis. Thus 224 OD pairs are considered because one OD pair has no demand for the considered period. The trip lengths and PT travel times are estimated using the average of those values at the IRIS level weighted by the demand. 
The considered road network, along with the regions and the boundaries forming the 15 origins and destinations, is to be found in Fig.~\ref{fig:network}. 

\begin{figure}
\begin{subfigure}{.49\textwidth}
  \centering
  \includegraphics[width=.95\linewidth]{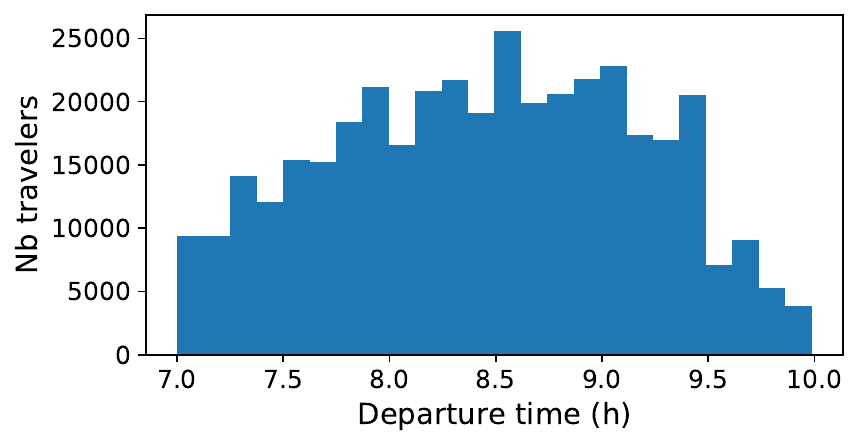}  
  \caption{}
\end{subfigure}
\begin{subfigure}{.49\textwidth}
  \centering
  \includegraphics[width=.95\linewidth]{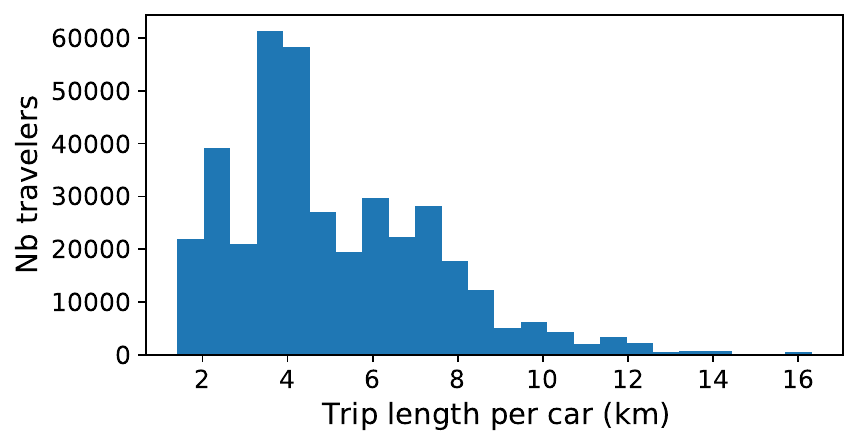}  
  \caption{}
\end{subfigure}
\newline
\begin{subfigure}{\textwidth}
  \centering
  \includegraphics[width=.5\linewidth]{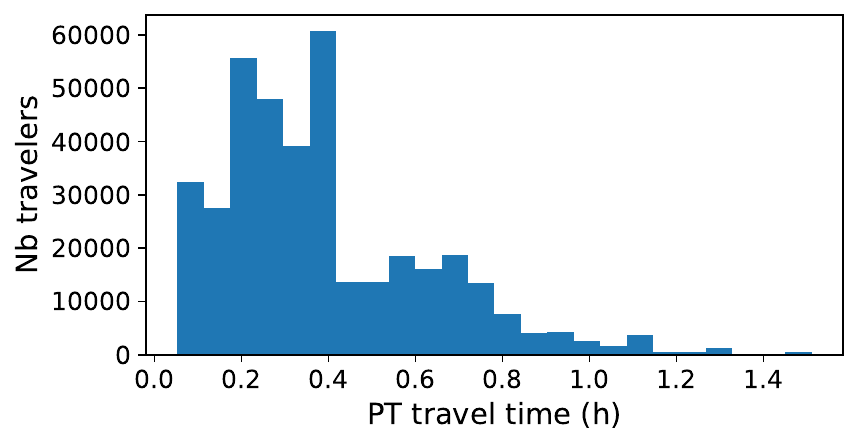}  
  \caption{}
\end{subfigure}
\caption{(a) Departure times, (b) trip lengths, and (c) PT travel times distributions.}
\label{fig:scenario_SC}
\end{figure}

A scenario is developed to test the proposed methodology. We consider the demand between 7:00 and 10:00 and split it into 15 minute subperiods (\cite{MLIDRM_2021}). Each period has its own PT travel time obtained from the navigator HERE and demand level per OD pair. 
The PT travel times for the trip from and to Lyon Metropolis are obtained using the HERE API (\cite{HERE_API}). For every subperiod and OD pair, the PT travel time is retrieved by sending a request to the navigator. The data from the navigator HERE considers the historical traffic conditions for each PT trip at a given hour of the day. Regarding the PT travel times for trips originating or ending outside of Lyon Metropolis, an average PT speed of 3~m/s (10.8~km/h) is used. This value is chosen to match the mean PT speed obtained from the navigator while being slightly lower to account for the inconvenience of switching mode at the city border (Park+Ride).
The departure times are generated uniformly for each subperiod. This scenario has 384~200 trips (or travelers). We use heterogeneous groups to ensure we have a proper granularity both in trips and departure times. They are aggregated with a maximum of 250 travelers per group and a minimum of two groups per OD pair and per hour. Thus, 2~163 groups are generated.
The distributions of the departure times, trip lengths, and PT travel times are shown in Fig.~\ref{fig:scenario_SC}.
It can be seen that there are no overlapping of the PT travel times and the trip lengths, meaning that the attractiveness of the PT strongly depends on the OD pair.

The default parameters used for the simulation can be found in Table~\ref{tab:params}. The VoT is chosen based on the work of \cite{Fosgerau2007TheStudy}. The carbon price is based on the European Union Emission Trading Scheme (see \cite{ICAP}). The maximum allowed variations $\epsilon_x$ and $\epsilon_p$ are taken as the inverse of the current iteration index (See \ref{sec::app_threshold} for a comparison with constant values). Practically, this reduces the exploration space size at each iteration to narrow the search when we come close to the modal equilibrium. The iteration process stops once the cost function $J$ is below the desired precision $J_\text{Goal}$, i.e., when the modal equilibrium is reached.

\begin{table}[h]
\caption{The default parameters used for the simulation.}\label{tab:params}
\begin{tabular*}{\hsize}{@{\extracolsep{\fill}}lll@{}}
Parameter & Notation & Value\\
\hline
VoT & $\alpha$ &   10.8 EUR/h\\
Endowment & $\kappa$ & 100 credits \\
Credit charge & $\tau$ &  200 credits \\
Price weight & $\eta$ & 1 \\ 
Cost function goal & $J_\text{Goal}$ & $10^{-3}$\\ 
Initial price & $p(0)$ & 0.01 EUR/credit \\
Initial modal shares &$\mathbf{x_0}(0)$& $\mathbf{0}$\\
Logit parameter & $\theta$ & 1 1/EUR\\
Emission weight & $\gamma$ & 50\\
Carbon price & $P_\text{carbon}$ & 20 EUR/tonne\\
\end{tabular*}
\end{table}

\subsection{Preliminary analysis}

\begin{figure}
\begin{subfigure}{.49\textwidth}
  \centering
  \includegraphics[width=.95\linewidth]{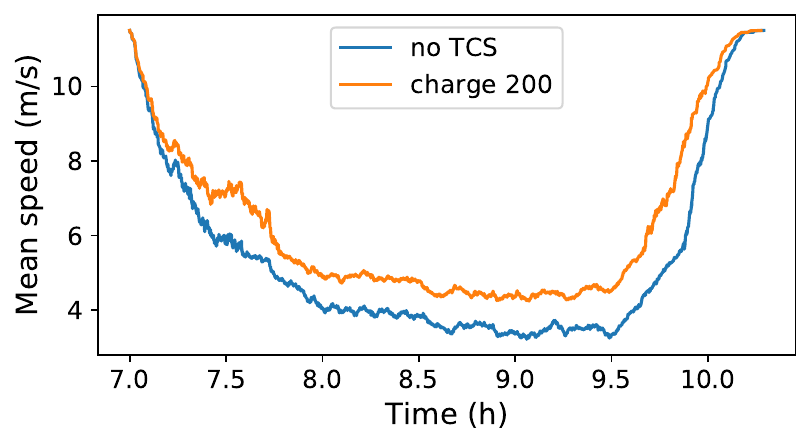}  
  \caption{}
\end{subfigure}
\begin{subfigure}{.49\textwidth}
  \centering
  \includegraphics[width=.95\linewidth]{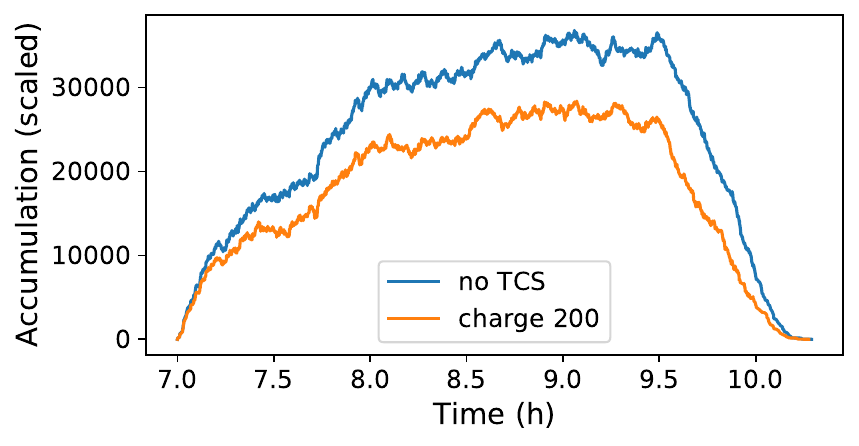}  
  \caption{}
\end{subfigure}
\newline
\begin{subfigure}{\textwidth}
  \centering
  \includegraphics[width=.5\linewidth]{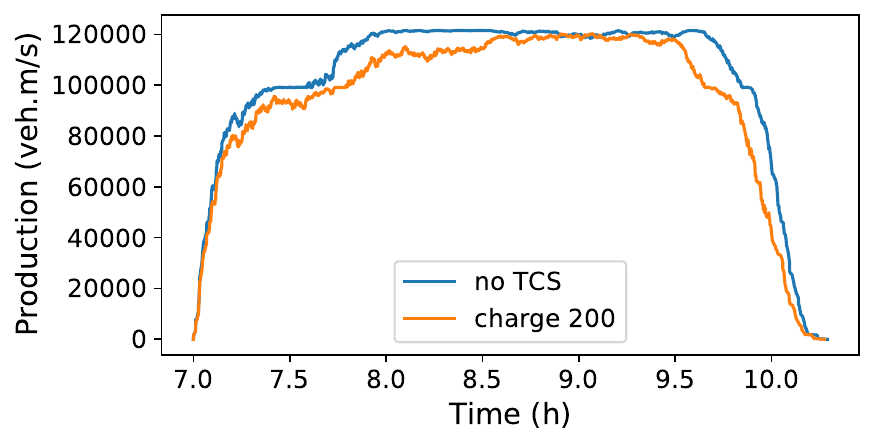}  
  \caption{}
\end{subfigure}
\caption{Speed (a), accumulation (b) and production (c) at the modal equilibrium for a credit charge of 200 credits and no TCS.}
\label{fig:MFD_SC2}
\end{figure}

First, we present the simulation results without and with TCS. It shows the congestion dynamics and helps to apprehend the scenario better.
The speed, accumulation, and production at the modal equilibrium along the simulation time are to be found in Fig.~\ref{fig:MFD_SC2}. The production is the product of the mean speed and the accumulation. It is the distance traveled by all the vehicles in the network per unit of time.

The traffic does not enter the hyper-congested regime, as the production does not decrease because of high accumulation. Under hyper-congestion, the PT alternative would be highly attractive, and thus, the car shares would decrease. Nevertheless, it undergoes clear loading, congested, and unloading stages. It permits to demonstrate the method capabilities for a realistic peak hour scenario.

\begin{figure}
\begin{subfigure}{.49\textwidth}
  \centering
  \includegraphics[width=.95\linewidth]{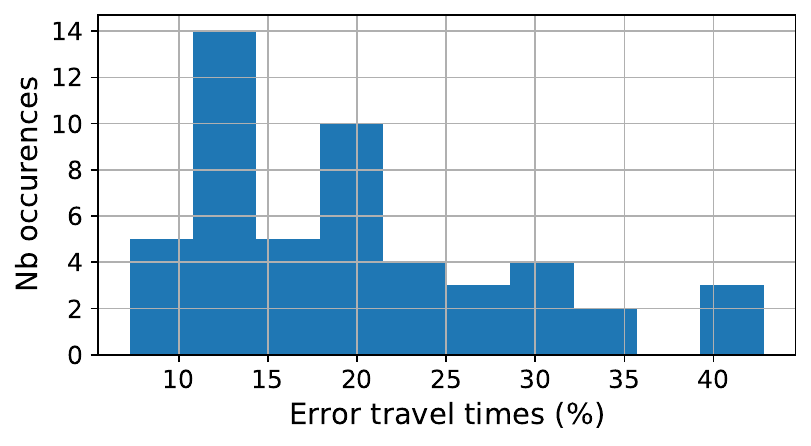}  
  \caption{}
\end{subfigure}
\begin{subfigure}{.49\textwidth}
  \centering
  \includegraphics[width=.95\linewidth]{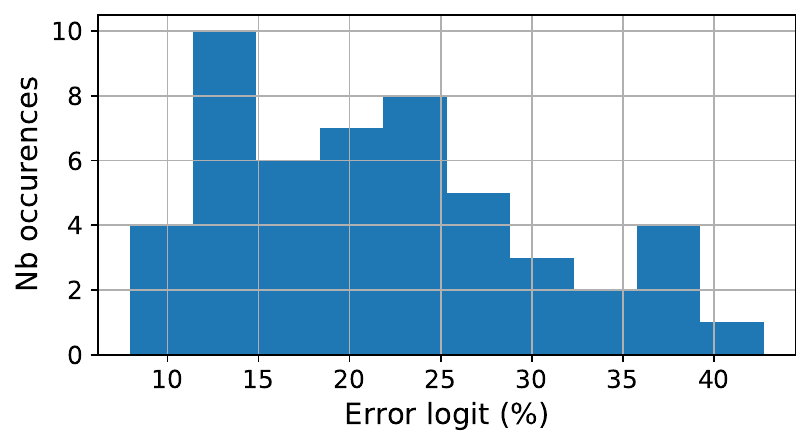}  
  \caption{}
\end{subfigure}
\caption{(a) Error on the travel times, and (b) error on the modal decisions.}
\label{fig:error_TT_logit_SC}
\end{figure}

Second, before investigating in details the equilibrium process, we assess the errors made by the linearization of the travel times. 50 pairs of modal shares $(\mathbf{x_0}, \mathbf{x_1})$ are randomly and separately generated following a uniform distribution. Simulations are carried out to define the exact values of $\mathbf{T(x_1)}$ and $\mathbf{T(x_0)}$. Then, the travel times and modal decisions are linearized around $\mathbf{x_0}$. Their values are approximated at $\mathbf{x_1}$ with the linearization. The norm of the error is normalized using the norm of the differences: $\lVert \mathbf{T(x_1)}-\mathbf{T(x_0)} \lVert_2$ and $\lVert \mathbf{\psi(x_1)}-\mathbf{\psi(x_0)} \lVert_2$. The results are presented in Fig.~\ref{fig:error_TT_logit_SC}.
The error of the linearization of the travel time and logit is lower than 45\%, with most of the occurrences below 25\%. It is satisfying as $\mathbf{x_1}$ is not always in the neighborhood of $\mathbf{x_0}$.

\subsection{Results}


\subsubsection{Comparing methods for computing equilibrium}

We directly feed the MSA with the optimal price derived by the new method based on travel times linearization. We do this because this method can only derive the modal shares and not the equilibrium price. Thus using another price value may lead to a different equilibrium and prevent us from a fair benchmarking. Note that, as MSA fails to calculate equilibrium prices, it greatly reduces the potential of this method in practice.

\begin{figure}
\begin{subfigure}{.49\textwidth}
  \centering
  \includegraphics[width=.95\linewidth]{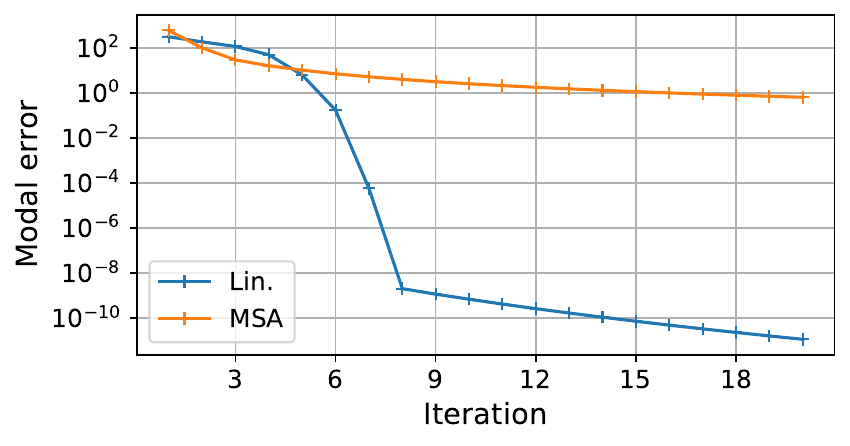}  
  \caption{}
\end{subfigure}
\begin{subfigure}{.49\textwidth}
  \centering
  \includegraphics[width=.95\linewidth]{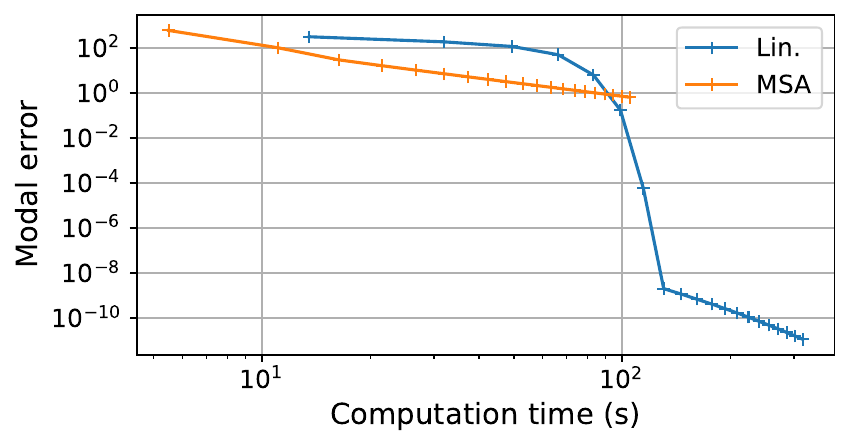}  
  \caption{}
\end{subfigure}
\newline
\begin{subfigure}{.49\textwidth}
  \centering
  \includegraphics[width=.95\linewidth]{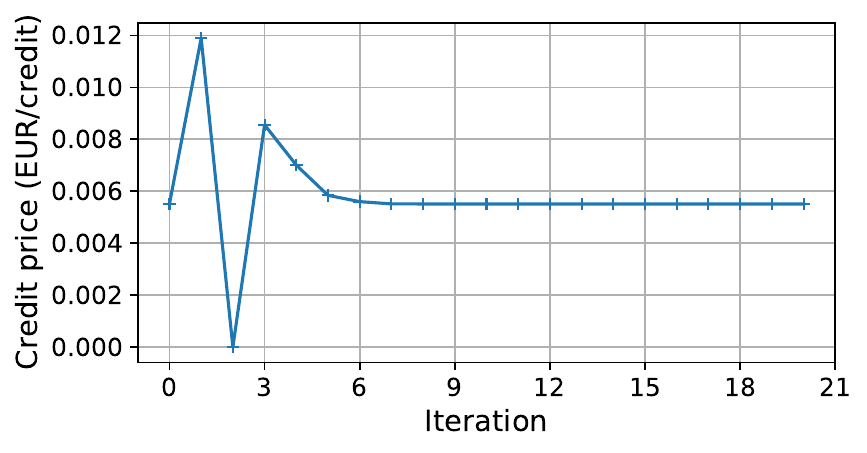}  
  \caption{}
\end{subfigure}
\begin{subfigure}{.49\textwidth}
  \centering
  \includegraphics[width=.95\linewidth]{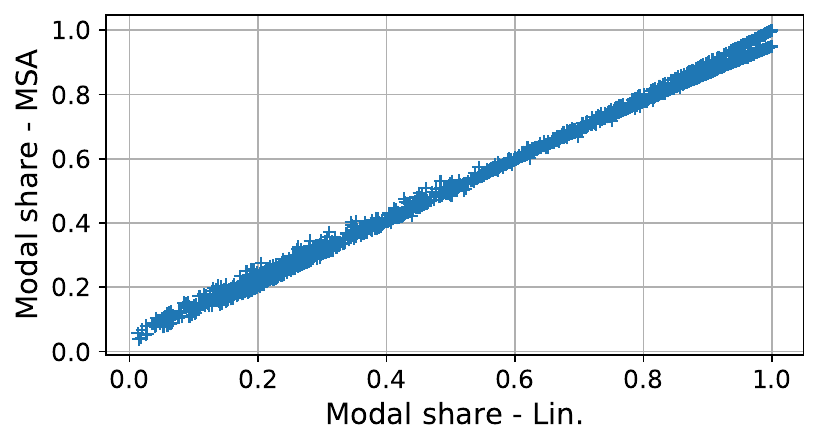}  
  \caption{}
\end{subfigure}
\caption{(a) Error between modal shares and decisions vs. iteration, (b) vs. computation time, (c) evolution of the credit price, and (d) modal shares at equilibrium.}
\label{fig:comp_SC2}
\end{figure}

Each method is run with 20 iterations. The modal errors $\frac{1}{2}\lVert \mathbf{x}-\mathbf{\psi} \lVert_2^2$, modal shares, and computation times are compared in Fig.~\ref{fig:comp_SC2} with an initial price of 0.00551~EUR/credit. 
The MSA is fast to compute but fails to reach high precision. The proposed methodology increases the computation burden by about one order of magnitude to increase the precision by about ten orders of magnitude.
Both methods found almost the same modal shares and credit price at equilibrium. The error between the modal share is only 4\%.

To further highlight the limits of the MSA, we run another equilibrium computation with another initial price of 0.001 EUR/credit. The equilibrium number of car users is then 217~695 with the MSA, which violates the credit cap as the limit is $\sum_{1}^N \gamma_i \kappa / \tau = 192~100$.

\subsubsection{Importance of departure times and trip lengths}

Most of the TCS frameworks proposed in the literature are based on Vickrey's bottleneck and BPR functions. They cannot account for the congestion dynamics and trip heterogeneity at the same time. We generate some alternative scenarios to highlight the importance of considering the heterogeneity in departure times and trip lengths. We show that the behavior of the TCS is greatly affected by a change in the departure time distribution or by the homogenization of the trips. 

\begin{figure}
\begin{subfigure}{.49\textwidth}
  \centering
  \includegraphics[width=.95\linewidth]{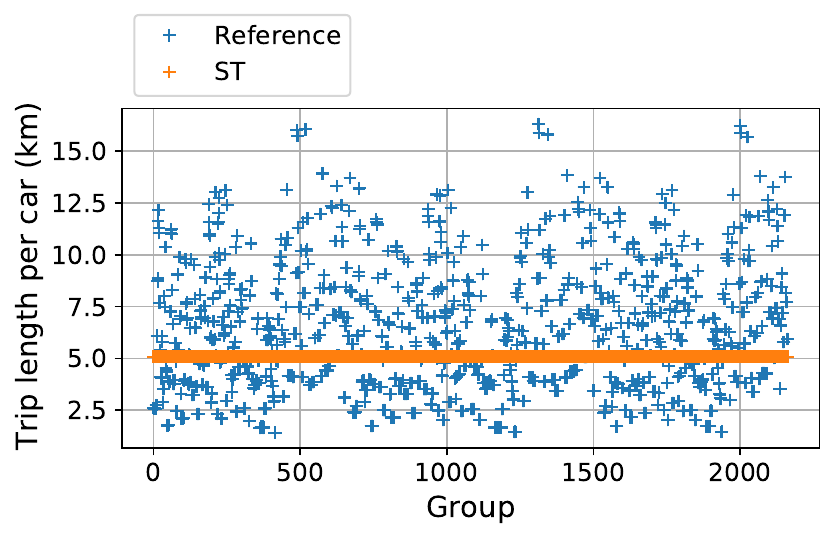}
  \caption{}
\end{subfigure}
\begin{subfigure}{.49\textwidth}
  \centering
  \includegraphics[width=.95\linewidth]{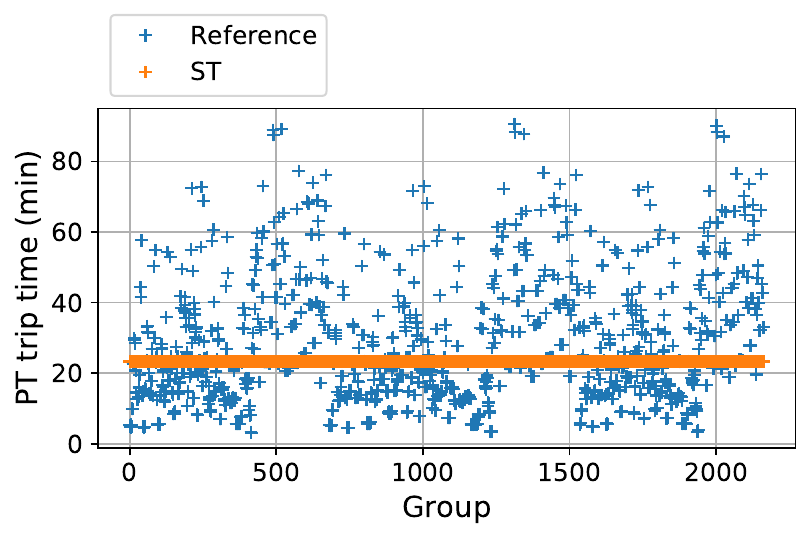}  
  \caption{}
\end{subfigure}
\newline
\begin{subfigure}{\textwidth}
  \centering
  \includegraphics[width=.6\linewidth]{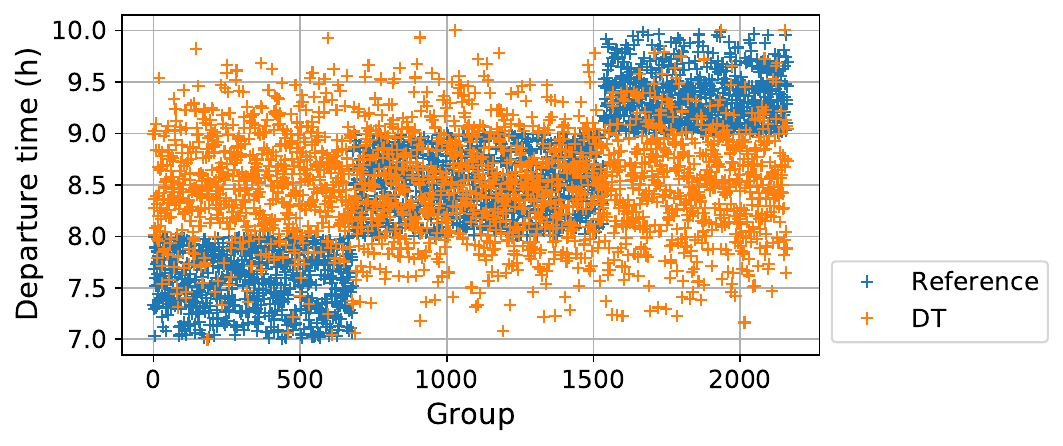}  
  \caption{}
\end{subfigure}
\caption{(a) Trip lengths and (b) PT travel times for ST and (c) departure times for DT.}
\label{fig:scenario_ST_DT_SC2}
\end{figure}

As Vickrey's bottleneck assumes that every traveler has the same trip, we create a scenario named ST, where all travelers have the same trip length and PT travel time. These parameters are computed by averaging the trip lengths and PT travel times weighted by the demand. 
As the BPR function does not consider the departure times, we create a scenario named DT where the departure times are generated with a different distribution. In the reference scenario, the departure times follow the distribution given in Fig.~\ref{fig:scenario_SC}. In scenario DT, they follow the normal distribution of mean 5400~s and standard deviation 1800~s. See Fig.~\ref{fig:scenario_ST_DT_SC2} for the differences between the reference scenario, ST, and DT.

\begin{figure}[h]
\centerline{\includegraphics[width=0.6\textwidth]{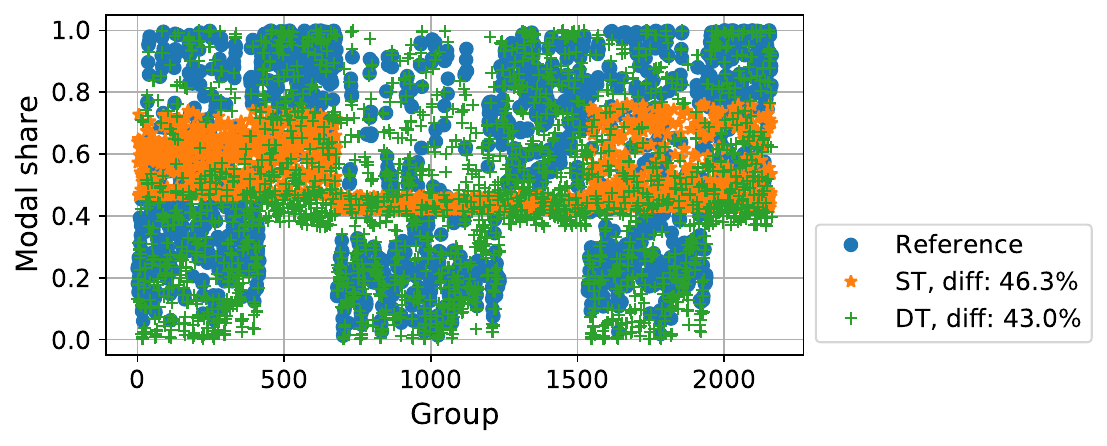}}
\caption{Modal shares at the equilibrium for the three scenarios with SC2.}
\label{fig:equil_ST_DT_SC2}
\end{figure}

The corresponding credit prices and modal shares at equilibrium are compared in Table~\ref{tab:prices_ST_DT_SC2} and Fig.~\ref{fig:equil_ST_DT_SC2}.
\begin{table}[h]
\caption{The credit prices and differences in modal shares at equilibrium for the three scenarios with the demand SC2.}\label{tab:prices_ST_DT_SC2}
\begin{tabular*}{\hsize}{@{\extracolsep{\fill}}llcc@{}}
Scenario & Price (EUR/credit) & Difference price & Difference modal shares\\
\hline
Reference & 0.00551 &   - & -\\
ST & 0.00820 & +48.8\% & 46.3\% \\
DT & 0 &  -100\% & 43.0\% \\
\end{tabular*}
\end{table}
With a more concentrated distribution of the departure times in DT, the traffic is significantly more congested. A credit charge of 200 credits is not a constraint anymore, as even without TCS, the PT is more attractive than the car. Thus the credits in DT do not have any monetary value. The difference of the modal shares is more than 40\%. 
Neglecting the congestion dynamics and assuming homogeneity of the trips leads to significant errors in estimating the modal shares at equilibrium. This simulation proves the necessity to consider both the heterogeneity in trip lengths and departure times.

\subsubsection{Sensitivity analysis}

%

\begin{figure}
\begin{subfigure}{.49\textwidth}
  \centering
  \includegraphics[width=.95\linewidth]{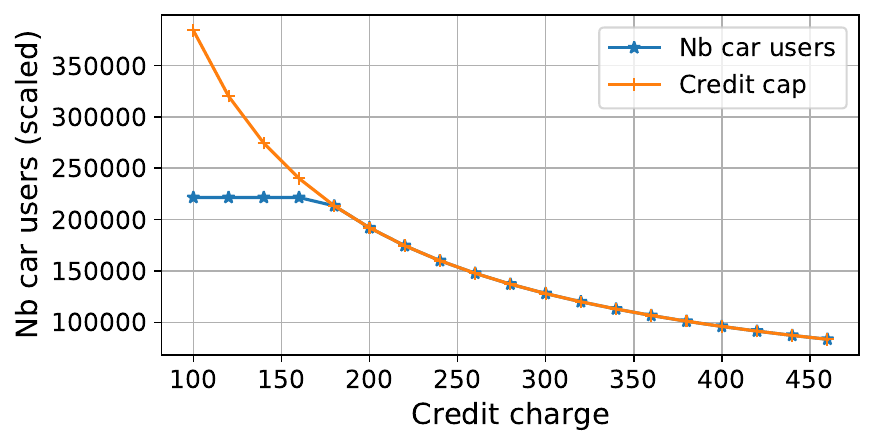}  
  \caption{}
\end{subfigure}
\begin{subfigure}{.49\textwidth}
  \centering
  \includegraphics[width=.95\linewidth]{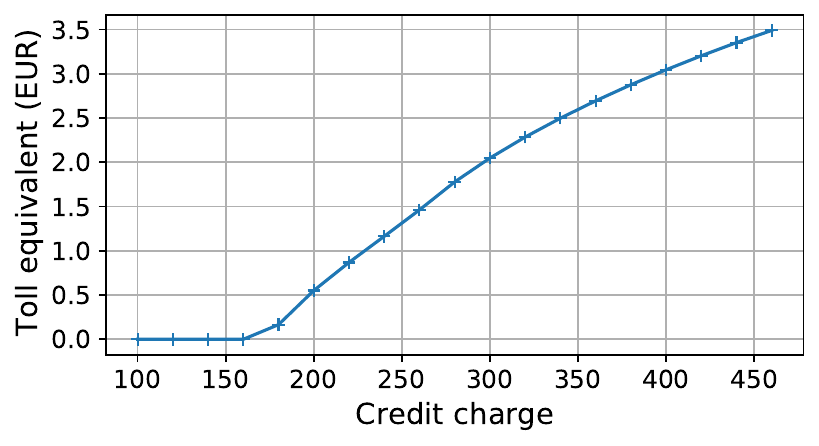}  
  \caption{}
\end{subfigure}
\newline
\begin{subfigure}{\textwidth}
  \centering
  \includegraphics[width=.45\linewidth]{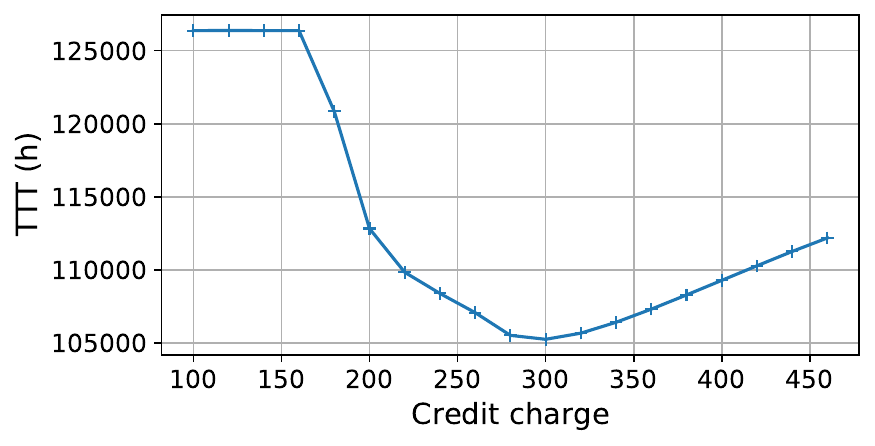}  
  \caption{}
\end{subfigure}
\caption{(a) Number of car users, (b) toll price in EUR, and (c) total travel time for different credit charges.}
\label{fig:sens_toll_SC2}
\end{figure}

Different credit charges are investigated to assess the impact of different TCS on the transportation system. The equilibriums are computed for credit charges between 100 and 460 credits with a step size of 20 credits. The number of car users and the toll equivalent $p (\tau-\kappa)$, i.e., the money a group has to spend to purchase the credits (on top of its allocation, which is for free) needed to take its car, are presented in Fig.~\ref{fig:sens_toll_SC2} for the different credit charges.
The TCS is only active from a credit charge of about 180 credits. Before, it does not constraints anyone on switching from car to PT. It can be seen that the price is zero when the credit cap is not constraining. It is in line with the MCC. As expected, the toll equivalent increases with the credit charge. It is expected: by augmenting the credit charge, the number of cars allowed on the network is reduced, and the ability to drive a car, here seen as a commodity, becomes scarce and thus more expensive. For a credit charge of 460 credits, which means less than one-quarter of the users can drive their private cars, the toll equivalent is around 3.5~EUR. Such a price is reasonable. For comparison, a transit ticket costs about 2~EUR in Lyon Metropolis as of 2021.
The evolution of the total travel time combines the increase of travel times for users switching from car to PT and the decrease caused by better traffic conditions for those still traveling by car. There seems to be a minimum for the total travel time at around 300 credits.

\begin{figure}
\begin{subfigure}{.49\textwidth}
  \centering
  \includegraphics[width=.95\linewidth]{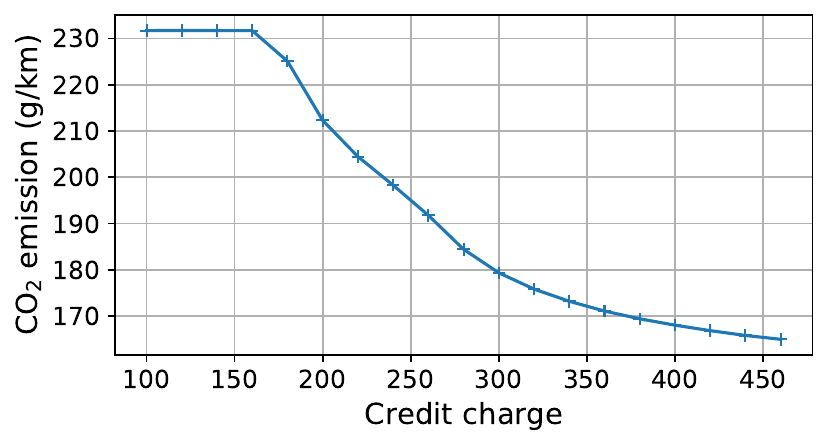}  
  \caption{}
\end{subfigure}
\begin{subfigure}{.49\textwidth}
  \centering
  \includegraphics[width=.95\linewidth]{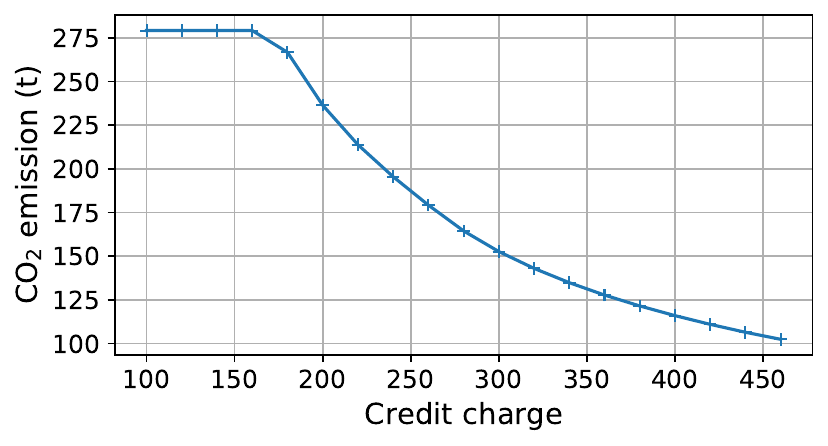} 
  \caption{}
\end{subfigure}
\caption{(a) \ensuremath{\mathrm{CO_2}} emission per distance and (b) \ensuremath{\mathrm{CO_2}} total emission for different credit charges.}
\label{fig:sens_toll_emi_SC2}
\end{figure}

The impact of the TCS on network carbon emission is also investigated in Fig.~\ref{fig:sens_toll_emi_SC2}.
The emission per distance decreases with the credit charge, as the lower accumulation permits better traffic conditions and a more efficient operating of the internal combustion engines. The total network emission decreases even more as the improvement of the performance of the combustion engines is coupled with a diminution of the number of cars on the network, i.e., the total traveled distance. A credit charge of 340 credits cuts the total network carbon emission by two.

\begin{figure}
  \centering
  \includegraphics[width=.49\linewidth]{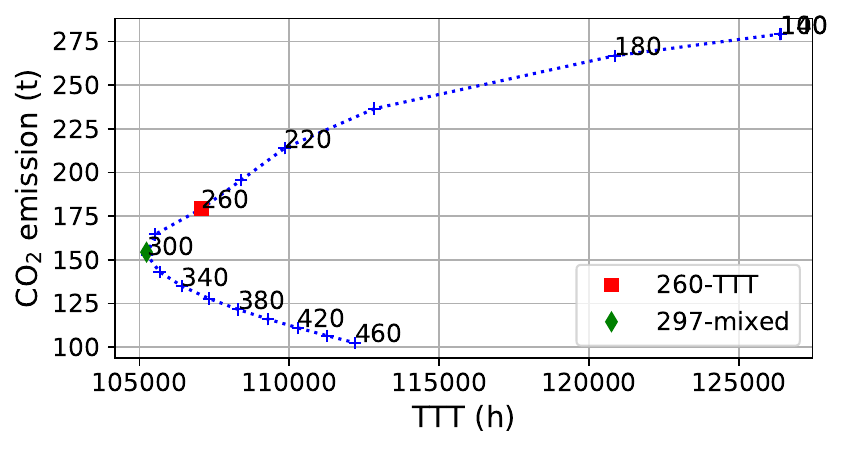} 
\caption{Total travel time vs. \ensuremath{\mathrm{CO_2}} emission for different credit charges. The green and red points are found by minimizing total travel time and the mixed objective function.}
\label{fig:sens_toll_ttt_emi_SC2}
\end{figure}

In Fig.~\ref{fig:sens_toll_ttt_emi_SC2} , we investigate the trade-off between total travel time and carbon emission.
The Pareto front for minimizing simultaneously total travel time and carbon emission, i.e., the set of non-dominated solutions, starts at a credit charge of about 300 credits.

\subsubsection{Optimize the credit charge}

\begin{figure}
\centering
\begin{subfigure}{\textwidth}
  \centering
  \includegraphics[width=.7\linewidth]{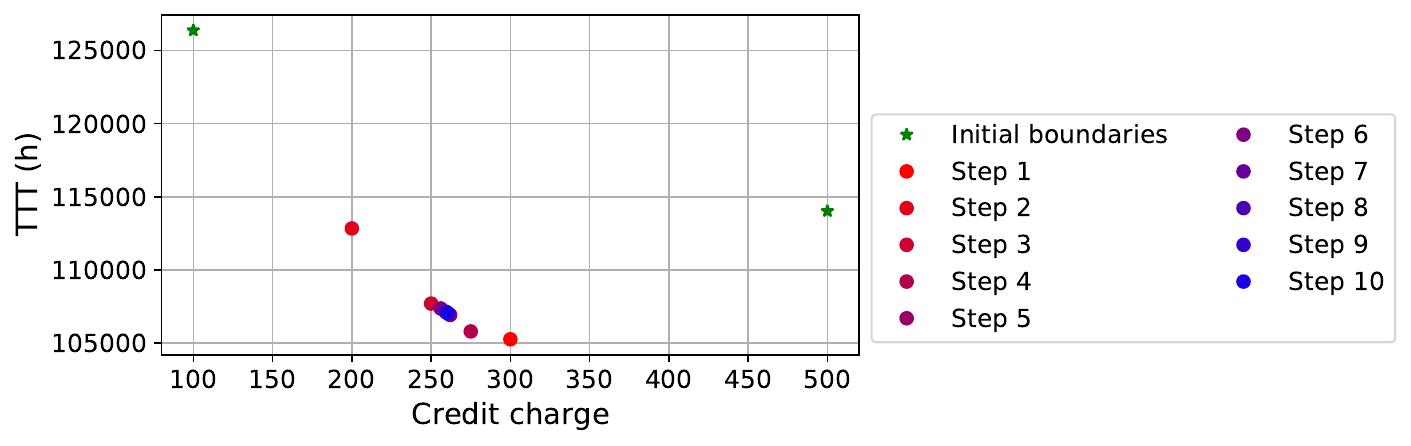}  
  \caption{}
\end{subfigure}
\newline
\centering
\begin{subfigure}{\textwidth}
  \centering
  \includegraphics[width=.7\linewidth]{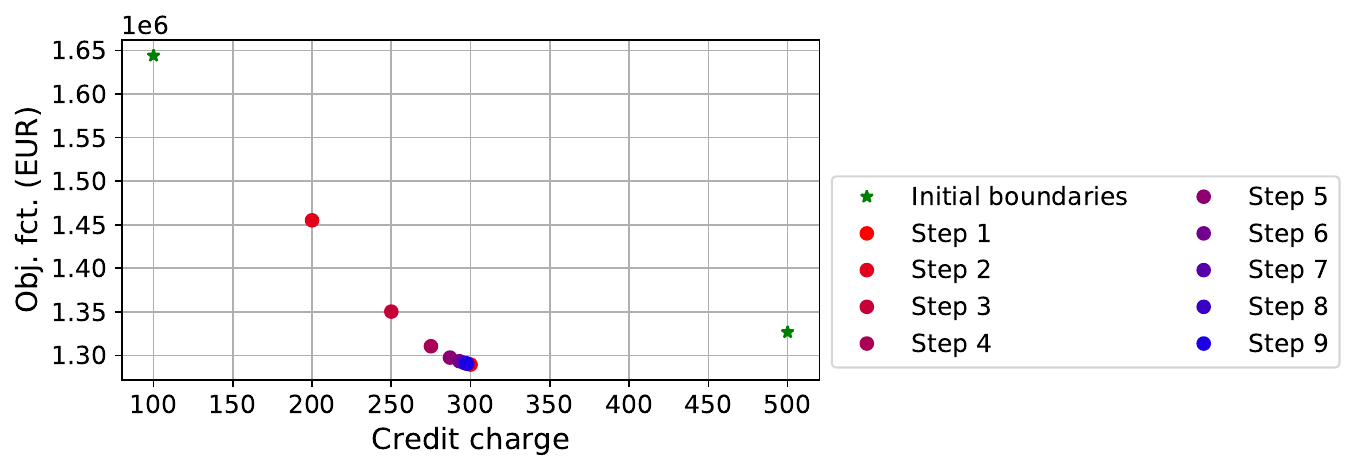} 
  \caption{}
\end{subfigure}
\caption{(a) Total travel time and (b) mixed objective function optimizations.}
\label{fig:toll_dichotomy_SC2}
\end{figure}

The credit charge optimization process by dichotomy is launched with an initial higher bound of 500 credits and an initial lower bound of 100 credits. The convergence of the process can be found in Fig.~\ref{fig:toll_dichotomy_SC2} for minimizing the total travel time only and the mixed objective function.

By trying to minimize the total travel time, the optimization process finds a credit charge of 260 credits, corresponding to a total travel time of 107~082~h. The optimal credit charge is actually 295 credits for a total travel time of 105~237~h. The error is only 2\%. It decreases the total travel time by 15\% by increasing the PT share by 20 points from 42\% to 62\%.
The optimization process with the mixed objective ends with a charge of 297 credits for a social cost of 1~290~878~EUR. The actual optimal credit charge is 330 credits for a social cost of 1~283~803~EUR. The proposed method found a value for the social cost 0.2\% away from the optimum in only nine iterations. To put it into perspective, using a greedy method and testing every credit charge between 100 and 500 credits would require 400 iterations, which means increasing the computation time by one to two orders of magnitude. Although the difference between the found and the optimal credit charge is relatively large, the difference with the objective function is minimal because the function is flat around the optimum. As expected, the credit charge found by minimizing the mixed objective is higher than the one minimizing the total travel time. It decreases the carbon emission by 45\% and the total travel time by 17\% by decreasing the car share by 24 points.

The total travel time and carbon emission are compared in Table~\ref{tab:ttt_emi_SC2} for the credit charges found by minimizing the total travel time (260 credits) and the mixed objective function (297 credits).
\begin{table}[h]
\caption{Total travel time and carbon emission with the two objective functions.}\label{tab:ttt_emi_SC2}
\begin{tabular*}{\hsize}{@{\extracolsep{\fill}}lccc@{}}
Objective & Total travel time (h) & Carbon emission (t)\\
\hline
No TCS & 126~369 & 279.2 \\
Total travel time & 107~082 & 179.3\\
Mixed objective & 105~239 & 154.3\\
\end{tabular*}
\end{table}
When minimizing the total travel time, the total travel time and the carbon emission are higher than when minimizing the mixed objective. We would expect the total travel time to be lower. By looking at those operating points in Fig.~\ref{fig:sens_toll_ttt_emi_SC2}, the credit charge of 260~credits found by minimizing the total travel time is not part of the Pareto front. However, relative to the total travel time without TCS, the error stays small.

\begin{figure}
\begin{subfigure}{.49\textwidth}
  \centering
  \includegraphics[width=.95\linewidth]{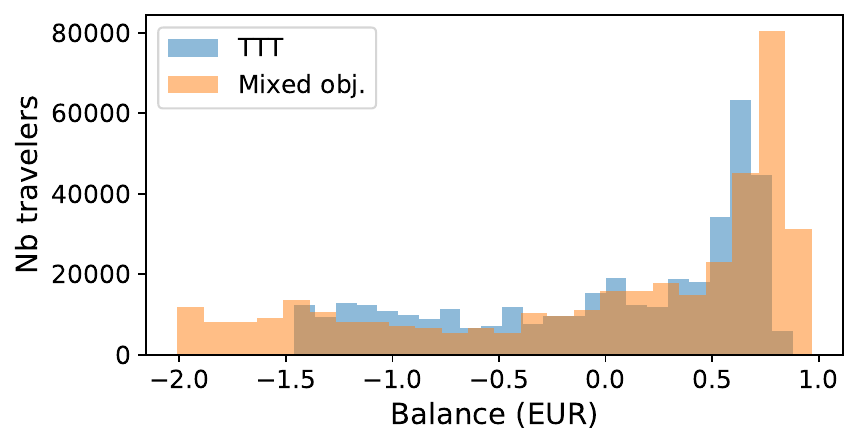}  
  \caption{}
\end{subfigure}
\begin{subfigure}{.49\textwidth}
  \centering
  \includegraphics[width=.95\linewidth]{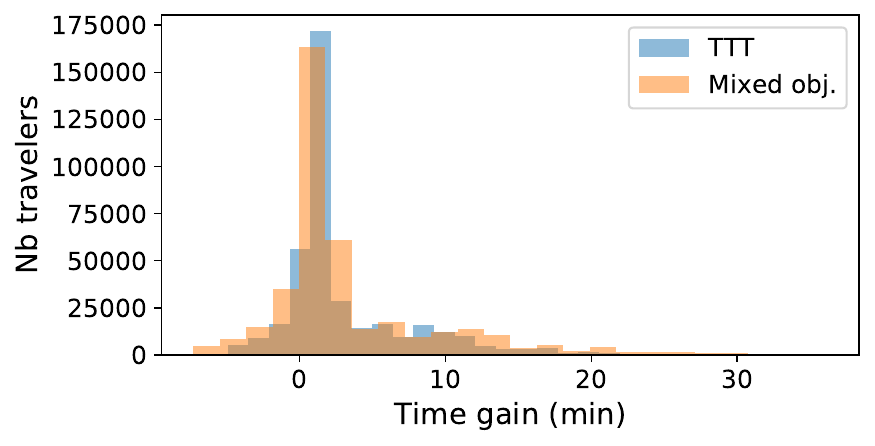} 
  \caption{}
\end{subfigure}
\newline
\begin{subfigure}{\textwidth}
  \centering
  \includegraphics[width=.5\linewidth]{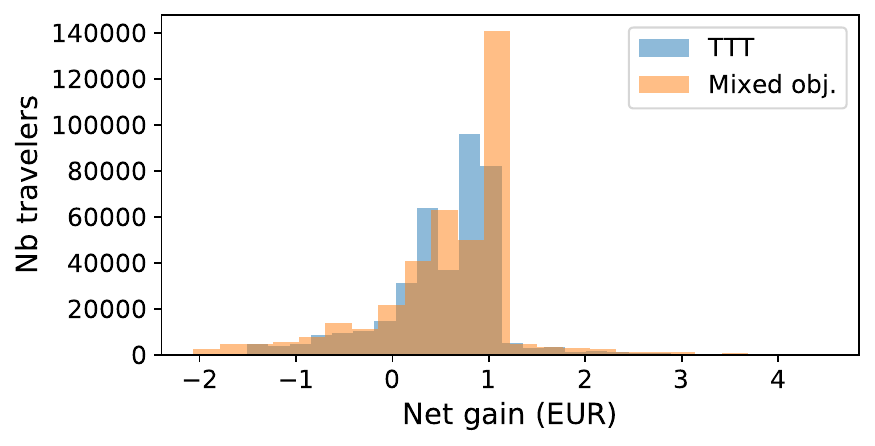} 
  \caption{}
\end{subfigure}
\caption{(a) Trade balances, (b) time gains, and (c) net gains for the credit charges found by minimizing the total travel time and the mixed objective.}
\label{fig:gains_SC2}
\end{figure}

We now look at the consequences for the different groups in Fig.~\ref{fig:gains_SC2} in terms of money earned with the credit trade:
\begin{equation}
    p(\kappa - x_i \tau),
\end{equation}
time gain:
\begin{equation}
    x_{i|\text{no TCS}} T_{i|\text{no TCS}} + (1-x_{i|\text{no TCS}}) T_i^\text{PT} - \left( x_{i} T_{i} + (1-x_{i}) T_i^\text{PT} \right)
\end{equation}
and net gain composed of the money balance from the trade of credits plus the change in travel times:
\begin{equation}
    p(\kappa - x_i \tau) + \alpha \left(x_{i|\text{no TCS}} T_{i|\text{no TCS}} + (1-x_{i|\text{no TCS}}) T_i^\text{PT} - \left( x_{i} T_{i} + (1-x_{i}) T_i^\text{PT}\right) \right).
\end{equation}
Positive values for these three indicators are gains, which means the implementation of the TCS brings benefits (additional revenue, reduced travel time). On the opposite, negative values are losses, which means the group suffers from the TCS (additional expenditure, increased travel time).
The groups spend up to 2~EUR and earn up to 1~EUR with the credit trade. Most of the groups save travel times, and the TCS increases some travel times by at most only five minutes. When it comes to the net gain of the system, most groups are net winners with a gain up to 1~EUR. Few travelers are losing up to 2~EUR. 

\section{Conclusions}\label{sec::conclusion}

A TCS is proposed with a trip-based MFD framework. The users are assumed to have fixed departure times and routes. They can choose between driving their private car and paying a credit charge or riding the PT using a logit model. No assumptions are made about the credit price mechanism.
The limits of the models used in most of the TCS literature (Vickrey's bottleneck and BPR function) are shown in a numerical example as they cannot consider the dynamics of the congestion or the heterogeneity of the trips.

Using the knowledge about the trip-based MFD formulation, the proposed method linearizes the travel times with respect to the modal shares. The modal equilibrium is reached with fewer iterations and with greater precision than the classical MSA, which cannot determine the equilibrium price value in the first place. Furthermore, the developed method ensures the credit cap is respected, and the credit price is found along with the modal shares.
The credit charge is optimized to find a trade-off between total travel time and carbon emission in a dichotomy-based approach.
A scenario based on the network and the demand of Lyon Metropolis are presented to illustrate the TCS and the methods to compute the travel times gradient, modal equilibrium, and optimal credit charge. Depending on the chosen objective function, the optimized credit charge decreases the total travel time by about 17\% and the carbon emission by about 45\%. Minimizing a mixed objective of total travel time and network emission results in a higher credit charge than minimizing the total travel time alone.
By considering the distribution of the gains, we highlight that the TCS benefits most of the users. However, there is still a minority for which the travel costs are increasing with the TCS. In this work, the credits are uniformly allocated. Especially, the allocation does not consider the heterogeneity of the OD pairs. Future work should consider leveraging the credit allocation to make the TCS profitable for as many users as possible.

The linearization of travel times presents more potential as it permits us to estimate changes in the neighborhood of a computed solution without resolving trip-based MFD for each candidate. Some other applications are worth to be investigated to assess the change of different measures (travel cost, pollution) with respect to policy instruments (traffic lights management, congestion pricing, TCS) or users' behaviors (modal choice, departure time, route).

\section*{Acknowledgments}

This project has received funding from the European Union's Horizon 2020 research and innovation program under Grant Agreement no. 953783 (DIT4TraM).

\section*{Authors contribution statement}

LB: Conceptualization; Data curation; Formal analysis; Investigation; Methodology; Visualization; Writing - original draft; Writing - review \& editing.

LL: Conceptualization; Funding acquisition; Methodology; Project administration; Supervision; Roles/Writing - original draft; Writing - review \& editing.

\section*{Conflicts of interest}
None.

\end{sloppypar} 

\typeout{}
\bibliography{references}

\appendix

\setcounter{table}{0}
\setcounter{figure}{0}

\section{Notations}\label{sec::app_notations}

Parameters, variables, and other notations are respectively summed up in Tables \ref{tab:summary_notations_parameters}, \ref{tab:summary_notations_variables}, and \ref{tab:summary_notations_others}.

\begin{table}[H]
\caption{Summary of parameters notations.}\label{tab:summary_notations_parameters}
\begin{tabular*}{\hsize}{@{\extracolsep{\fill}}ll@{}}
Notation & Meaning\\
\hline
$\alpha$ & VoT\\
$\gamma_i$ & Number of travelers in group $i$\\
$\Gamma$ & \ensuremath{\mathrm{CO_2}} weight for the credit charge optimization \\ 
$\kappa$ & credit allocation \\
$\tau$ &  credit charge \\
$\eta$ & price weight for the QP \\ 
$\theta$ & logit parameter \\ 
$C_{i}^\text{PT}$ & travel cost of group $i$ by PT\\
$T_{i}^\text{PT}$ & travel time per PT of group $i$ \\
$l_i$ & trip length of group $i$\\
$t_i$ & departure time of group $i$ \\
$P_\text{carbon}$ & carbon price \\
\\
\end{tabular*}
\end{table}

\begin{table}[H]
\caption{Summary of variables notations.}\label{tab:summary_notations_variables}
\begin{tabular*}{\hsize}{@{\extracolsep{\fill}}ll@{}}
Notation & Meaning\\
\hline
$c$ & local slope of the gradient of the speed (taken as positive)\\
$C_{i}^\text{car}$ & travel cost of group $i$ by driving its car \\
$e, g$ & event: either the entry or the exit of an group on the network\\
$e_{i,e}$ & event when group $i$ ends its trip\\
$e_{i,s}$ & event when group $i$ starts its trip\\
$E$ & total network \ensuremath{\mathrm{CO_2}} emission\\
$E_\text{dist}$ & \ensuremath{\mathrm{CO_2}} emission per distance\\
$f(t)$ & distance traveled by the virtual traveler until time $t$\\ 
$i, j$ & index of an group, which represents a group of travelers\\ 
$l_e$ & distance traveled between the events $e$ and $e+1$\\
$L_m$ & mean traveled distance by car\\
$L_m^w$ & mean traveled distance by car weighted by the absolute values of the gradient of the logit\\
$L_\text{tot}$ & total traveled distance \\
$n$ & accumulation at a given time \\
$\bar{n}$ & typical accumulation\\
$N_c$ & number of car users\\
$p$ & credit price\\
$t_e$ & time at which the event $e$ occurs \\
$T_e$ & time between the events $e$ and $e+1$ \\
$T_i$ & travel time per car of group $i$ \\
$T_\text{dept}$ & departure time window\\
$TT_\text{c}$ & mean travel time per car \\
$TT_\text{c}^w$ &  mean travel time per car weighted by the absolute values of the gradient of the logit\\
$TT_\text{PT}^w$ &  mean travel time per PT weighted by the absolute values of the gradient of the logit\\
$TTT$ & total travel time \\
$V$ & mean speed in the network at a given time \\
$\bar{V}$ & mean speed in the network over the whole simulation \\ 
$w_i$ & absolute value of the gradient of the logit\\
$\mathbf{x}$ & shares of groups taking the car\\
$\mathbf{\tilde x}$ & concatenation of modal shares and credit price\\
$\mathbf{\psi}$ & modal decisions of the groups\\
\\
\end{tabular*}
\end{table}

\begin{table}[H]
\caption{Other notations.}\label{tab:summary_notations_others}
\begin{tabular*}{\hsize}{@{\extracolsep{\fill}}ll@{}}
Notation & Meaning\\
\hline
$\cdot_0$ & starting value\\
$\Delta \cdot$ & difference of the value of the variable compared to its reference\\
$\nabla \cdot$ & gradient of the variable related to the modal shares \\
$\tilde \nabla \cdot$ & gradient of the variable related to the modal shares and the credit price \\
\\
\end{tabular*}
\end{table}

\section{Equilibrium derivation with multi-modal MFD}\label{sec::app_3D}

In a complete multi-modal congestion model, the speeds of the PT vehicles and of the cars depend on the accumulation of the PT and of the cars. The proposed methodology can be extended to this case, and the main changes are presented below.

In the derivation the logit decision model in Eq.~(\ref{eq::logit_grad}), the variation of the PT travel time should be considered:
\begin{equation}\label{eq::logit_grad_3D}
    \begin{cases}
        \psi_{0,i} &= \frac{e^{- \theta (\alpha T_{0,i} + \tau p_0) }}{e^{- \theta (\alpha T_{0,i} + \tau p_0) }+ e^{- \theta \alpha T_{0,i,\text{PT}}}};\\
        \tilde \nabla \psi_{i,j} &= \psi_{0,i} (\psi_{0,i}-1) \theta \alpha \nabla T_{i,j} + \psi_{0,i} (1-\psi_{0,i}) \theta \alpha \nabla T_{i,j}^\text{PT};\\
        \tilde \nabla \psi_{i,N+1} &= \psi_{0,i} (\psi_{0,i}-1) \theta \tau.
    \end{cases} 
\end{equation}
It can be seen that the car ratio augments with the PT travel time, which is expected.

The gradient of the speed Eq.~(\ref{eq::grad_speed}) is extended to:
\begin{equation}\label{eq::grad_speed_3D}
        \nabla V^m_{e,i} =  \gamma_i \delta_i(e) \frac{\partial V^m}{\partial n}(n_{e-1}, n_{e-1}^\text{PT}) + \frac{1}{C_\text{PT}} \gamma_i \delta_i^\text{PT}(e) \frac{\partial V^m}{\partial n_\text{PT}}(n_{e-1}, n_{e-1}^\text{PT}),
\end{equation}
for the mode $m$ being the car or the PT. $\delta_i(e)=1$ if the users of group $i$ are in the network between $e-1$ and $e$ with their cars and 0 otherwise. $\delta_i^\text{PT}(e)=1$ if the PT alternative of group $i$ is in the network between $e-1$ and $e$ and 0 otherwise. $C_\text{PT}$ is a correction factor to link the number of travellers riding PT to the accumulation of buses. Indeed a bus serves several travelers and part of the travelers will ride subway or tramway, without impacting the traffic conditions on the road network.

The case distinctions to compute the gradient of inter-event times $\nabla T_e$ need to account for events linked to entry and exits of buses (i.e. PT riders) as well. The entry times of buses are the same as the cars' ones, but the exit times are different since the travel times are different. Eq.~(\ref{eq::grad_entry_entry}), (\ref{eq::grad_exit_entry}), and (\ref{eq::grad_all_exit}) still hold.

When accounting for additional modes (bikes, scooters), it becomes necessary to introduce another mode ratio per group and increase the size of the problem, but the methodology still holds.

\section{Numerical evaluation for the condition of uniqueness}\label{sec::app_uni}

\begin{figure}
  \centering
  \includegraphics[width=.5\linewidth]{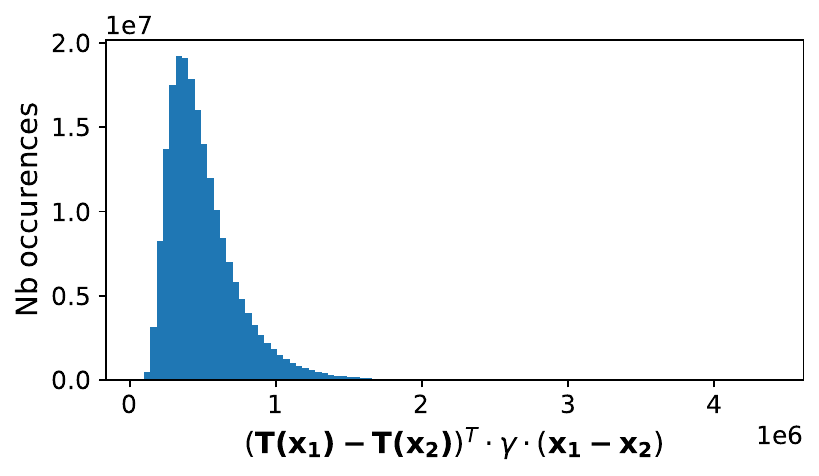}
\caption{Computation of the dot product of the car travel time differences and weighted modal share differences.}
\label{fig:monotony_tt_w_x_check}
\end{figure}

We evaluate numerically the assumption Eq.~(\ref{eq::monotony_tt_w_x}) for the numerical use case in Sect.~\ref{sec::example}. 20~000 modal share vectors $\mathbf{x}$ are generated using a Latin Hypercube sampling. It represents about ten points per dimension. The Eq.~(\ref{eq::monotony_tt_w_x}) is computed for every pair of different points. The distribution of the dot product is represented in Fig.~\ref{fig:monotony_tt_w_x_check}.
The assumption Eq.~(\ref{eq::monotony_tt_w_x}) seems to hold for our numerical example. The equilibrium might be unique.

\section{Stability analysis of the equilibrium}\label{sec::app_stab}

In this study, we do not consider a day-to-day adjustment process where the users learn from previous days to define their mode choices and their credit buying strategy. We instead develop a semi-analytical framework to define the equilibrium state of the considered transportation system directly. Investigating the stability of such an equilibrium requires introducing a time-dependent (day-to-day in practice) process reproducing the evolution of mode choice and credit price when deviations from the equilibrium are observed.

As we look for deviations from the equilibrium, we can assume the credit price is non-zero. The credit price is then adjusted based on the difference between credit offer (allocation) and demand (credit charge times modal decision). Mode shares adapt with time following the differences between modes shares and the logit-based decisions associated with the actual mode share vector. In other words, the dynamics process tends to match the fixed-point observed at the equilibrium. The two following equations give the system time dynamics:
\begin{equation}\label{eq::time_evol}
\begin{cases}
    \mathbf{\dot{x}} &= \psi(\mathbf{x}, p)-\mathbf{x}; \\
    \dot{p} &= \sum_{i=1}^N \gamma_i (\tau \psi_i(\mathbf{x}, p) - \kappa).
\end{cases}
\end{equation}
Around the equilibrium $\mathbf{\tilde x^*}$, the linearization gives an estimation of the dynamics and the time derivative of $\mathbf{\Delta \tilde x} = \mathbf{\tilde x}-\mathbf{\tilde x^*}$ is approximated by 
\begin{equation}
    \mathbf{\dot{\Delta \tilde x}} = \mathbf{A} \mathbf{\Delta \tilde x},
\end{equation}
with the Jacobian $\mathbf{A}$ defined by
\begin{equation}
\begin{cases}
    A_{i,j} = \nabla \psi_{i,j} - \delta_{i,j} \ \forall \ i \in [1,N], j \in [1,N+1]; \\
    A_{N+1,j} = \tau \sum_{i=1}^N \gamma_i \nabla \psi_{i,j} \ \forall \ j \in [1,N]; \\
    A_{N+1,N+1} = \tau \sum_{i=1}^N \gamma_i \nabla \psi_{i,N+1}, \\
\end{cases}
\end{equation}
with $\delta_{i,j}=1$ if and only if $i=j$ and 0 otherwise.

Stability is ensured when the real parts of the eigenvalues are all strictly negative (see Theorem 4.7 in \cite{Khalil2002}, knowing the chosen evolution function defined by Eq.~(\ref{eq::time_evol}) is continuously differentiable). While we cannot verify that such a condition holds for any equilibrium state, because the system dynamics $\mathbf{x} \mapsto \mathbf{T}(\mathbf{x})$ and thus the logit decision-making $(\mathbf{x},p) \mapsto \mathbf{\psi}(\mathbf{x},p)$ has no explicit formulation. However, it is straightforward to check if this condition holds when the equilibrium state values have been derived. We calculated the eigenvalues of the Jacobian for all credit charges between 100 and 460 credits with a step size of 20 credits considering the equilibrium values for $\mathbf{x^*}$ and $p^*$ for our numerical test case (see \hyperref[sec::example]{Sect.~\ref*{sec::example}}).
In all cases, the real parts of the eigenvalues are all strictly negative, and the equilibrium is then asymptotically stable with respect to modal shares and credit price.

\setcounter{figure}{0}
\section{Sensitivity of the threshold for the search space}\label{sec::app_threshold}

\begin{figure}
\begin{subfigure}{.49\textwidth}
  \centering
  \includegraphics[width=.95\linewidth]{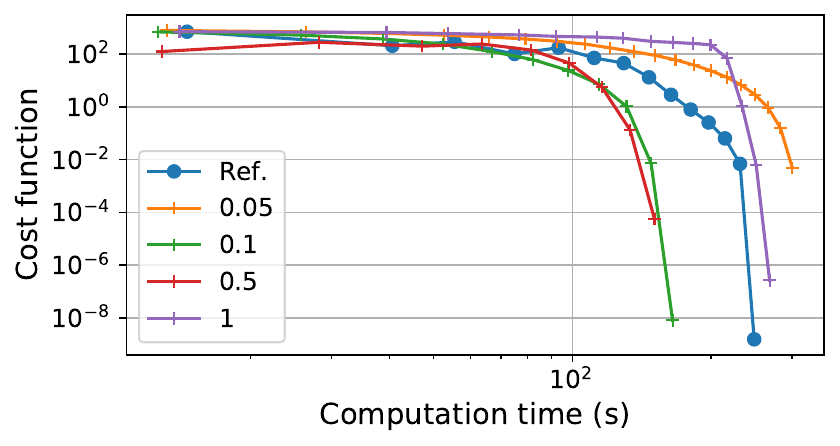}
  \caption{}
\end{subfigure}
\begin{subfigure}{.49\textwidth}
  \centering
  \includegraphics[width=.95\linewidth]{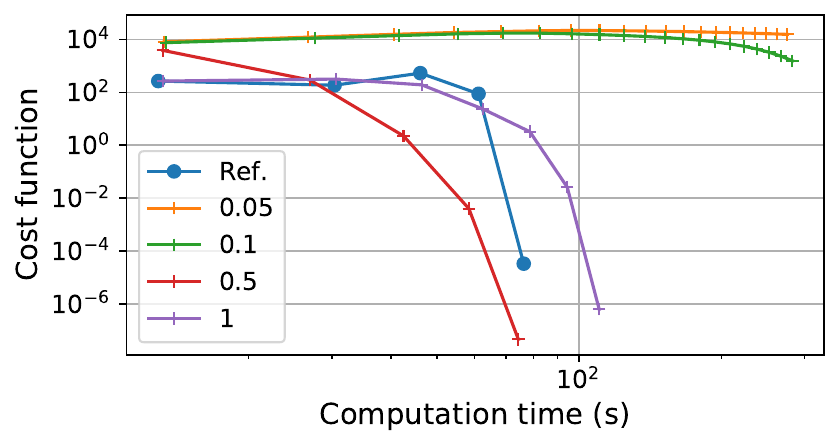}  
  \caption{}
\end{subfigure}
\newline
\begin{subfigure}{\textwidth}
  \centering
  \includegraphics[width=.5\linewidth]{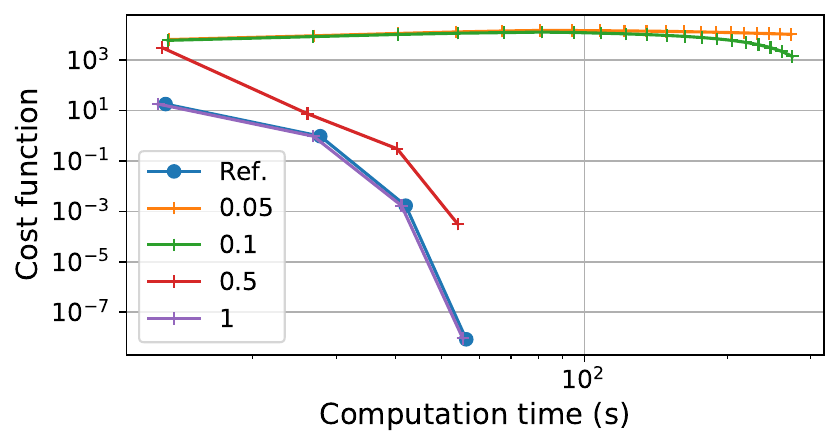}  
  \caption{}
\end{subfigure}
\caption{Cost function values vs. computation time for different maximum allowed variations with (a) no TCS; (b) a credit charge of 200 credits; and (c) a credit charge of 300 credits.}
\label{fig:ann_thres}
\end{figure}

To assess the sensitivity of the search for a modal equilibrium with regard to the maximum allowed variations $\epsilon_x$ and $\epsilon_p$, different constant thresholds 0.01, 0.05, 0.1, 0.5, and 1 are compared to the inverse of the time step (Ref.) in Fig~\ref{fig:ann_thres} for no TCS, credit charge of 200, and 300 credits.

Setting the allowed maximum variation too low makes the convergence more difficult. When convergence occurs, all the values lead to the same equilibrium. There is no best value for the maximum allowed variations in terms of computation time. The chosen approach with the inverse of the step size is a good compromise.

\end{document}